\documentclass[12pt, american]{article}%
\usepackage{amssymb}
\usepackage{amsfonts}
\usepackage{amsmath}
\usepackage{lmodern}
\usepackage[nohead]{geometry}
\usepackage[doublespacing]{setspace}
\usepackage[bottom]{footmisc}
\usepackage{booktabs} 
\usepackage{endnotes}
\usepackage{natbib}
\usepackage{graphicx,caption,subfigure}
\usepackage{rotating}
\usepackage{hyperref}
\usepackage[latin1]{inputenc} 
\usepackage[T1]{fontenc} 
\usepackage{pdflscape}
\usepackage{rotating}
\usepackage[flushleft]{threeparttable}
\usepackage{dsfont}
\usepackage{dcolumn}
\newcolumntype{.}{D{.}{.}{4}}
\usepackage{%
  babel,     
  fixltx2e,  
}
\newenvironment{proof}[1][Proof]{\noindent\textbf{#1.} }{\ \rule{0.5em}{0.5em}}
\makeatletter
\def\@biblabel#1{\hspace*{-\labelsep}}
\makeatother
\hypersetup{colorlinks, linkcolor=blue, citecolor=blue, urlcolor=blue}

\usepackage{amsthm}
\theoremstyle{plain}
\newtheorem{assumption}{Assumption}

\newtheorem{lemma}{Lemma}
\theoremstyle{definition}
\newtheorem{proofoflemma}{Proof of Lemma}

\newtheorem{theorem}{Theorem}
\theoremstyle{definition}
\newtheorem{proofof}{Proof of Theorem}

\theoremstyle{definition}

\theoremstyle{definition}

\newtheorem{remark}{Remark}

\begin{document}

\title{Oracle Efficient Estimation of Structural Breaks\\
 in Cointegrating Regressions}
\author{Karsten Schweikert\thanks{Address: University of Hohenheim, Core Facility Hohenheim \& Institute of Economics, Schloss Hohenheim 1 C, 70593 Stuttgart, Germany, e-mail: \textit{karsten.schweikert@uni-hohenheim.de}}
\medskip\\
\date{\normalsize [Latest update: \today]}}

\maketitle

\sloppy

\singlespacing

\begin{abstract}

\noindent

\footnotesize
In this paper, we propose an adaptive group lasso procedure to efficiently estimate structural breaks in cointegrating regressions. It is well-known that the group lasso estimator is not simultaneously estimation consistent and model selection consistent in structural break settings. Hence, we use a first step group lasso estimation of a diverging number of breakpoint candidates to produce weights for a second adaptive group lasso estimation. We prove that parameter changes are estimated consistently by group lasso and show that the number of estimated breaks is greater than the true number but still sufficiently close to it. Then, we use these results and prove that the adaptive group lasso has oracle properties if weights are obtained from our first step estimation. Simulation results show that the proposed estimator delivers the expected results. An economic application to the long-run US money demand function demonstrates the practical importance of this methodology.

\strut


\end{abstract}

\thispagestyle{empty}

\pagebreak%
\onehalfspacing

\section{Introduction}\label{sec:intro}
\noindent


In this paper, we consider modelling cointegration relationships where the long-run equilibrium may differ for subsamples, thereby allowing for (multiple) structural breaks in the cointegrating regression. We assume that cointegration holds over some (fairly long) period of time, but then shifts to a new `long-run' relationship. The number of breaks and their location are unknown to the researcher. Although coefficients of long-run equilibrium equations are relatively persistent by definition, accounting for the possibility of structural breaks is crucial in cointegration analysis, which usually involves long sample periods. On the one hand, long time series are needed to study the long-run behaviour of economic systems, on the other hand, employing long time series increases the likelihood of encountering structural change during the sample period. It is widely known that structural breaks, when present, can mask cointegrating relationships and render cointegration tests uninformative \citep{Campos1996, GregoryNasonWatt1996, Qu2007}. Hence, we propose a two-step approach to detect (multiple) structural breaks in cointegrating regressions using penalized regression techniques.

Since time series used for economic analyses have become very long in some instances, detecting (multiple) structural breaks has emerged as an important problem in the econometrics literature. For comprehensive surveys on structural breaks in time series models (`change-point' detection in the statistics literature or `pattern recognition' in the context of signal processing), see, for example, \cite{Perron2006}, \cite{AueHorvath2013} and \cite{NiuHaoZhang2015}. While classical structural break models for linear regressions attempt to detect one unknown break via a grid search procedure \citep{Andrews1993}, it is not feasible to use grid searches for the detection of multiple breaks because the computational cost increases exponentially with the presumed number of breaks (needing least squares operations of order $O(T^m)$ for $m$ breaks). Addressing this issue, \cite{BaiPerron1998, BaiPerron2003} use dynamic programming techniques (henceforth Bai-Perron algorithm), requiring at most least-squares operations of order $O(T^2)$ for any number of breaks, to add breaks sequentially to the model. Recently, several approaches have been proposed that reframe the task of detecting and estimating structural breaks as a model selection problem employing penalized regressions and related model selection techniques \citep{Davis2006, Harchaoui2010, JinShiWu2013, ChanYauZhang2014, Ciuperca2014, JinWuShi2016, QianJia2016, QianSu2016, Behrendt2020}. Instead of grid search procedures which augment linear regression models with parameter changes, model selection procedures take a top-down approach and try to shrink the set of all possible breakpoint candidates to contain only the true breakpoints. These approaches benefit from high computational efficiency and detect structural breaks with high accuracy.

The theory for (multiple) structural breaks in cointegrating regressions is not nearly as developed as the theory for change-points in the statistics and signal processing literature. Most studies are concerned with cointegration testing in the presence of structural instability. One of the most popular cointegration tests with an unknown breakpoint is the one proposed by \cite{GregoryHansen1996, GregoryHansen1996b} in which the location of the break can be estimated via grid search at the minimum of the individual cointegration test statistics. \cite{Hatemi-J2008} extends the test to account for two breaks and \cite{Schweikert2019} allows for the possibility of nonlinear adjustment to the long-run equilibrium. \cite{Maki2012} employs a hybrid procedure, detecting $m-1$ breaks by minimizing the sum of squared residuals among all possible sample splits and finally determining the last break by minimizing a cointegration test statistic. Unfortunately, these tests are non-informative about the location of breaks. They optimize the model specification to provide evidence against the null hypothesis of no cointegration, thereby not necessarily finding those breakpoints which optimize the model fit. \cite{Maki2012} determines the first $m-1$ breaks based on improving the model fit but does not do so for the last breakpoint. Hence, the set of estimated breakpoints is not completely informative. Other studies with a strong focus on cointegration testing which conduct breakpoint estimation as a by-product are \cite{Carrion-i-Silvestre2006} and \cite{AraiKurozumi2007}. They propose a CUSUM-based approach to test the null hypothesis of cointegration with a structural break against the alternative hypothesis of no cointegration. \cite{Qu2007} considers a cointegrated system allowing the cointegrating rank to change during different subsamples so that it is possible to detect cointegrating relationships that exist only in some subsamples. \cite{WesterlundEdgerton2006} design LM-based test statistics invariant to structural breaks to test the null of no cointegration and \cite{DavidsonMonticini2010} use subsample procedures to account for structural breaks in their cointegration tests. 

In contrast, few studies are primarily focussed on modelling structural change in cointegrated systems. \cite{KejriwalPerron2008, KejriwalPerron2010} propose to estimate the number and location of structural breaks in cointegrating equations by applying the Bai-Perron algorithm. Inference on breakpoints is studied in, among others, \cite{BaiLumsdaineStock1998}, \cite{QuPerron2007}, \cite{KejriwalPerron2008, KejriwalPerron2010}, \cite{LiPerron2017} and \cite{OkaPerron2018}. Using penalized regression approaches to account for structural breaks in cointegrating regression has not been explored yet in great detail. A similar idea has been proposed by \cite{SchmidtSchweikert2019} but their procedure is limited to bivariate cointegrating regressions using a modified adaptive lasso estimator. Here, we extend their methodology to cointegrating regressions with multiple regressors and provide a rigorous proof that the adaptive group lasso estimator is oracle efficient in settings with an unknown number of breaks and a diverging number of breakpoint candidates.

The proposed estimation method in this paper consists of two main steps: in the first step, we apply the group lasso estimator to a cointegration model with a diverging number of breakpoint candidates. We allow that breaks can occur at any point in time except for some lateral trimming which is mostly needed to identify the baseline coefficients in the first regime. We prove that the group lasso estimator consistently estimates parameter changes. However, it is well-known that lasso estimators are not simultaneously parameter estimation consistent and model selection consistent in situations where the restricted eigenvalue condition or related conditions such as the strong irrepresentable condition do not hold \citep{ChanYauZhang2014}. Under these conditions, we show that the number of selected breaks is greater than the true number of breaks almost surely, but their estimated location is sufficiently close to their true location. In the second step, we use the first step group lasso estimates as weights for the adaptive group lasso. We provide a rigorous proof that the adaptive group lasso has the oracle properties if the first step algorithm assumes a maximum number of breaks and the distance between breaks depends on the sample size ensuring that the breakpoint candidates for the second step estimation are sufficiently distinct. The number of breaks is then estimated as the number of non-zero groups obtained after adaptive group lasso optimization. 

The paper is organized as follows. \autoref{sec:method} describes the proposed adaptive group lasso procedure to estimate structural breaks in cointegrating regressions. \autoref{sec:sim} is devoted to the Monte Carlo simulation study. \autoref{sec:app} reports the results of an empirical application of our methodology to the US money demand function, and \autoref{sec:conc} concludes. Proofs of all theorems in the paper are provided in the Mathematical Appendix.

\section{Methodology}\label{sec:method}
\noindent
In the following, we specify a cointegrated system with multiple structural breaks at which it attains new equilibrium states. The cointegrated system does not deviate persistently from each equilibrium until the next break occurs and a new equilibrium is maintained.

\subsection{Framework}\label{sec:framework}
\noindent
Let $\lbrace y_t \rbrace^{\infty}_{t=1}$ denote a scalar process generated by
\begin{equation}
y_t = \sum\limits_{j=1}^{m+1} \left[ \mu + \boldsymbol{\beta}'_j \boldsymbol{X}_t + u_t \right] \mathds{1}{\{t_{j-1} \leq t < t_j \}} , \qquad t = 1, 2, \dots,
\label{eq:ci}
\end{equation}
where $t_j, j \in \lbrace 0, 1, \dots m + 1 \rbrace$ denote the breakpoints $1 = t_0 < t_1 < \dots < t_{m+1} = T + 1$. $\mu$ is the intercept, $\boldsymbol{\beta}'_j = (\beta_{j1}, \beta_{j2}, \dots, \beta_{jN})$ are regime-dependent coefficients and $\lbrace \boldsymbol{X}_t \rbrace^{\infty}_{t=1}$, where $\boldsymbol{X}_t = (X_{1t}, X_{2t}, \dots, X_{Nt})'$, follows an $N$-vector integrated process\footnote{Note that this specification of the process rules out integrated regressors with a deterministic drift component. Relaxing this assumption would be relatively straightforward.}
\begin{equation}
\boldsymbol{X}_t = \boldsymbol{X}_{t-1} + v_t, \qquad t = 1, 2, \dots,
\label{eq:rw}
\end{equation}
where $\boldsymbol{X}_0 = 0$. $\lbrace u_t \rbrace^{\infty}_{t=1}$ and $\lbrace v_t \rbrace^{\infty}_{t=1}$ are mean-zero weakly stationary error processes. For expositional simplicity, we restrict our analysis to cointegrating regressions with a constant intercept across regimes.\footnote{Our main results generally hold if the coefficients of included deterministic components, e.g. linear and quadratic trend terms, do not change over the sampling period. A discussion of breaks in the intercept is included in the supplementary material to this paper.} We make the following assumptions about the vector process $w_t = (u_t, v'_t)'$:
\begin{assumption}\label{as:1}
The vector process $\lbrace w_t \rbrace^{\infty}_{t=1}$ satisfies the following conditions:
\begin{enumerate}
\item[(i)] $E w_t = 0$ for $t = 1, 2, \dots$
\item[(ii)] $\lbrace w_t \rbrace^{\infty}_{t=1}$ is weakly stationary.
\item[(iii)] $\lbrace w_t \rbrace^{\infty}_{t=1}$ is strong mixing with mixing coefficients of size $-p \beta/(p - \beta)$ and $E |w_t|^{p} < \infty$ for some $p > \beta > 5/2$. 
\end{enumerate}
Further, we assume that the long-run covariance matrix $\Omega_{v} = \sum\limits_{j=-\infty}^{\infty} E v_t v_{t-j}'$ is positive definite.
In addition, we require that
\begin{equation*}
\underset{T}{\sup} E \left\vert \frac{1}{T} \sum\limits_{i=1}^{s} X_{ji} u_i \right\vert^{4+\epsilon} < \infty, \qquad \text{for } 1 \leq j \leq N, \, 1 \leq s \leq T \text{ and some } \epsilon > 0.
\end{equation*}
\end{assumption}
While the first three conditions of Assumption \ref{as:1} are standard in cointegration analysis, assuming that $\Omega_v$ is positive definite implies that $\boldsymbol{X}_t$ is non-cointegrated. We denote the number of structural breaks by $m$. While the number of true structural breaks $m_0$ is unknown, we assume that the maximum number of structural breaks $m^*$ is known to the researcher. The estimated number of breakpoints is denoted by $\hat{m}$. The locations of breakpoints relative to sample size, so-called break fractions, are denoted by $\tau_{j} = t_{j} / T, j \in \lbrace 0, 1, \dots m + 1 \rbrace$. 

Throughout this paper, we use the following notation to present our main results: let $y_T = (y_1, y_2, \dots, y_T)'$ denote the vector containing $T$ observations of our response variable and $u_T = (u_1, u_2, \dots, u_T)'$ denotes the error term vector. The vector of $T$ observations for the $N$-dimensional variable $\boldsymbol{X}_t$ is denoted by $\boldsymbol{X} = (\boldsymbol{X}_1, \dots, \boldsymbol{X}_T)'$. Our design matrix $\boldsymbol{Z}_T$ is an $T \times TN$ matrix defined by
\begin{equation}
\boldsymbol{Z}_T = \begin{pmatrix}
\boldsymbol{X}_1' & 0 & 0 & \dots & 0 \\
\boldsymbol{X}_2' & \boldsymbol{X}_2' & 0 & \dots & 0 \\
\boldsymbol{X}_3' & \boldsymbol{X}_3' & \boldsymbol{X}_3' & \dots & 0 \\
\vdots\\
\boldsymbol{X}_T' & \boldsymbol{X}_T' & \boldsymbol{X}_T' & \dots & \boldsymbol{X}_T' \\
\end{pmatrix},
\end{equation}
and we define the Gram matrix $\boldsymbol{\Sigma} = \boldsymbol{Z}_T' \boldsymbol{Z}_T / T^2$. Adjacent columns of $\boldsymbol{Z}_T$ differ only by one entry which means that the columns are almost identical for $T \to \infty$. Consequently, $\boldsymbol{\Sigma}$ does not converge to a positive definite asymptotic counterpart. It follows that the restricted eigenvalue condition \citep{BickelRitovTsybakov2009} does not hold and we cannot establish our consistency proofs based on this assumption. See \cite{ChanYauZhang2014} for a thorough discussion of this issue.

We set $\boldsymbol{\theta}_1 = \boldsymbol{\beta}_1$ and
\begin{equation}
\boldsymbol{\theta}_i = \begin{cases}
\boldsymbol{\beta}_{j+1} - \boldsymbol{\beta}_j, & \text{when $i = t_j$}, \\
\boldsymbol{0}, & \text{otherwise},
\end{cases}
\end{equation} 
for $i = 2, \dots, T$. For the remainder of this article, $\boldsymbol{\theta}_i = \boldsymbol{0}$ means that $\boldsymbol{\theta}_i$ has all entries equaling zero and $\boldsymbol{\theta}_i \neq \boldsymbol{0}$ means that $\boldsymbol{\theta}_i$ has at least one non-zero entry. The coefficient vector $\boldsymbol{\theta}(T) = (\boldsymbol{\theta}_1, \boldsymbol{\theta}_2, \dots, \boldsymbol{\theta}_T)'$ is of length $TN$ and contains all time-specific parameter changes. Because we treat structural breaks as rare events and assume that parameter changes persist for some time, the number of non-zero elements in $\boldsymbol{\theta}(T)$ is assumed to be small, i.e.\ smaller than $m^* + 1$ groups of size $N$.

We denote the true value of a parameter with a $0$ superscript. $\lbrace \tau^0_j, j = 1, \dots, m_0 \rbrace$ denotes the set of true break fractions and $\boldsymbol{\beta}^0_j$, $j = 1, \dots, m_0 + 1$ defines the true coefficient of the $j$-th regime. For technical reasons, we additionally set $\boldsymbol{\beta}^0_0 = 0$. We define the index sets $\bar{\mathcal{A}} = \lbrace 1 \leq i \leq T: \boldsymbol{\theta}^0_i \neq \boldsymbol{0} \rbrace$ denoting the indices of truly non-zero coefficients (including the baseline coefficient) and $\mathcal{A} = \lbrace i \geq 2: \boldsymbol{\theta}^0_i \neq \boldsymbol{0} \rbrace$ denoting the non-zero parameter changes. The index set obtained from our first step estimation belonging to all estimated non-zero parameter changes is denoted by $\mathcal{A}_T = \lbrace i \geq 2: \tilde{\boldsymbol{\theta}}_i \neq \boldsymbol{0} \rbrace$. We note that the first regime's coefficient (before the first breakpoint) is not allowed to be zero.\footnote{While the cointegrating vector $(1,\boldsymbol{0}')'$ in principle ensures that $y_t = \mu + u_t$ is stationary under our assumptions, we exclude this case to simplify our exposition. In the following, we need a clear distinction between zero and non-zero coefficients to decide whether their indices belong into the sets $\bar{\mathcal{A}}$ or $\bar{\mathcal{A}}^c$. Allowing zero baseline coefficients would require several case-by-case considerations.} Since we indicate breakpoints with non-zero coefficients in our penalized regression approach, the set $\mathcal{A} = \lbrace t^0_1, t^0_2, \dots, t^0_{m_0} \rbrace$ is also used to denote true breakpoints. Similarly, the set $\mathcal{A}_T = \lbrace \hat{t}_1, \hat{t}_2, \dots, \hat{t}_{\hat{m}} \rbrace$ denotes estimated breakpoints, i.e., indices of those coefficients which are estimated to be non-zero. $|\mathcal{A}|$ denotes the cardinality of the set $\mathcal{A}$ and $\mathcal{A}^c$ denotes the complementary set. We use those sets to index rows and columns of vectors and matrices. For example, let $\boldsymbol{Z}_{T, \mathcal{A}}$, $\boldsymbol{Z}_{T,\mathcal{A}^c}$ contain the columns of $\boldsymbol{Z}_T$ and $\boldsymbol{\theta}_{\mathcal{A}}(T)$, $\boldsymbol{\theta}_{\mathcal{A}^c}(T)$ contain the rows of $\boldsymbol{\theta}(T)$ associated with active and inactive breakpoints, respectively.

For notational convenience, we use `$\Rightarrow$' to signify weak convergence of the associated probability measures and $\overset{p}{\to}$ to denote convergence in probability. Continuous stochastic processes such as a Brownian motion $B(s)$ on [0,1] are simply written as $B$ if no confusion is caused. We also write integrals with respect to the Lebesgue measure such as $\int\limits_{0}^{1} B(s)ds$ simply as $\int\limits_{0}^{1} B$. Throughout the paper, several (distinct) large constants are all denoted with $C$, while small constants are denoted by $\epsilon$. 

Using these definitions, our cointegration model described in Equation~\eqref{eq:ci} can be expressed as a high-dimensional regression model in matrix form
\begin{equation}
y_T = \boldsymbol{Z}_T \boldsymbol{\theta}(T) + u_T.
\label{eq:ci.mat}
\end{equation}
Since only $m_0 + 1$ groups within $\boldsymbol{\theta}(T)$ are truly non-zero, we need to obtain a sparse solution to the high-dimensional regression problem in Equation~\eqref{eq:ci.mat}. This means we frame the detection of structural breaks as a model selection problem and use available methods from this strand of the literature. To reduce the dimensionality of the estimation problem, we assume that breaks occur for all coefficients simultaneously. This allows us to treat all regressors at each point in time as one group. We can therefore apply the group lasso estimator proposed by \cite{YuanLin2006} to achieve a sparse solution. As our first step, we minimize the objective function,
\begin{equation}
Q^*(\boldsymbol{\theta}(T)) = \frac{1}{T} \Vert y_T - \boldsymbol{Z}_T \boldsymbol{\theta}(T) \Vert^2 + \lambda_T \sum\limits_{i=1}^{T} \Vert \boldsymbol{\theta}_i \Vert,
\label{eq:gfl}
\end{equation}
to obtain the group lasso estimator for $\boldsymbol{\theta}(T)$ which is henceforth denoted by $\tilde{\boldsymbol{\theta}}(T) = \arg \min_{\boldsymbol{\theta}(T)} \, Q^*$. $\lambda_T$ is the tuning parameter and $\Vert \cdot \Vert$ denotes the $L_2$-norm. Unfortunately, the group lasso estimator inherits the same problems, namely estimation inefficiency and model selection inconsistency, as the plain lasso estimator. Similar to the idea first presented in \cite{Zou2006}, we reestimate the objective function with individual coefficient weights to alleviate this problem and to try to reduce the number of falsely detected breaks. The statistical properties of adaptive group lasso estimators for a fixed number of groups are investigated in \cite{WangLeng2008}. Since we have a diverging set of breakpoint candidates, least squares estimation of the full model is not feasible. However, we show that group lasso is a consistent estimator for non-zero parameter changes giving us appropriate weights for a second step adaptive group lasso estimation. This approach is similar to the ideas put forth in \cite{WeiHuang2010}, \cite{HorowitzHuang2013}, \cite{SchmidtSchweikert2019}, and \cite{Behrendt2020}. 

As will be demonstrated later, the group lasso estimator only slightly overselects breaks under the right tuning. The algorithm employed to estimate $\tilde{\boldsymbol{\theta}}(T)$ allows to pre-specify the maximum number of breakpoint candidates $M$, i.e.\ the maximum number of non-zero groups in $\tilde{\boldsymbol{\theta}}(T)$, and the minimum distance between breaks. Since the group lasso overselects breaks in the first step, $M$ should be set large enough to encompass all true breakpoints and some additional falsely selected non-zero groups. This condition guarantees that $\tilde{\boldsymbol{\theta}}(T)$ always contains $MN$ elements. In turn, $TN - MN$ columns of $\boldsymbol{Z}_T$ corresponding to zero coefficients are eliminated during the first step to result in the $T \times MN$ design matrix $\boldsymbol{Z}_S$. Hence, for given $M \ll T$, the column size of the new design matrix is substantially smaller than the original size $TN$ and does not longer depend on the sample size. This allows us to further assume that all eigenvalues of $\boldsymbol{\Sigma}_S = \boldsymbol{Z}_S' \boldsymbol{Z}_S / T^2$ are contained in the interval $[c_*, c^*]$, where $c_*$ and $c^*$ are two positive constants. This means that we can relate to a restricted eigenvalue condition similar to \cite{BickelRitovTsybakov2009} for the second step estimation. While the restricted eigenvalue condition in general does not hold for change-point settings, the dimension reduction of the first step allows us to postulate this assumption for our reduced design matrix. It should be noted that our assumption for the second step estimation is not restrictive for empirical applications because the notion of a long-run equilibrium relationship implies a maximum number of breaks and a minimum regime length. A minimum regime length is further justified by the minimum subsample size needed to precisely estimate parameter changes. Consequently, $M$ should be chosen so that the average regime length in case of equidistantly-spaced breaks still guarantees enough observations per regime to estimate all coefficient changes.

We follow \cite{WangLeng2008} and define the adaptive group lasso objective function
\begin{equation}
Q(\boldsymbol{\theta}_S) = \frac{1}{T} \Vert y_T - \boldsymbol{Z}_S \boldsymbol{\theta}_S \Vert^2 + \lambda_S \sum\limits_{i=1}^{M} w_i \Vert \boldsymbol{\theta}_{S,i} \Vert,
\label{eq:agfl}
\end{equation}
where $\gamma > 0$ and $w_i$ are the group-specific weights assigned as follows
\begin{equation*}
w_i =  \begin{cases}
\begin{array}{lll}
\Vert \tilde{\boldsymbol{\theta}}_{S,i} \Vert^{-\gamma} & \text{ if } & \tilde{\boldsymbol{\theta}}_{S,i} \neq 0 \\
\infty &  \text{ if } & \tilde{\boldsymbol{\theta}}_{S,i} = 0,
\end{array}
\end{cases}
\end{equation*}
and set $0 \times \infty = 0$. $\tilde{\boldsymbol{\theta}}_{S,i}$, $i = 1, \dots, |\mathcal{A}_T| + 1$ denotes the non-zero group lasso coefficient estimates obtained from optimizing the objective function in Equation~\eqref{eq:gfl}. The remaining $M - |\mathcal{A}_T| - 1$ group elements of $\tilde{\boldsymbol{\theta}}_S$ can be filled with zero groups as long as their selected indices lead to $\boldsymbol{\Sigma}_S$ being a positive definite matrix for all $T$.

We denote the estimator minimizing $Q(\boldsymbol{\theta}_S)$ with $\hat{\boldsymbol{\theta}}_S = \arg \min_{\boldsymbol{\theta}_S} \, Q$. The weight of the first coefficient is usually set to zero to ensure that the system is cointegrated with a cointegrating vector different from $(1,\boldsymbol{0}')'$ if no structural break occurs. Eliminating columns from the initial design matrix requires a mapping of our second step indices to recover the original indices. For notational convenience, we use the mapping $g: \mathbb{N} \to \mathbb{N}, \; i \mapsto g(i) = t_i$, where $t_i$ is the breakpoint corresponding to the index $i$, for this purpose and define the index set $\bar{\mathcal{A}}^*$ ($\mathcal{A}^*$) to pick out the elements that correspond to truly non-zero coefficients (parameter changes).

We note that the major computing cost comes from the first step group lasso estimation considering a large number of observations as potential breakpoints. The second step represents a marginal addition to the total computing time if the first step estimation was sufficiently successful in eliminating inactive breakpoint candidates. The interested reader may consult \cite{ChanYauZhang2014} for a detailed discussion of computational complexity in this context.\footnote{While the Bai-Perron algorithm needs at most $O(T^2)$ operations, the group LARS algorithm used to solve Equation~\eqref{eq:gfl} has computational burden of order $O(M^3 + MT)$. Hence, the group LARS algorithm has a stronger dependence on the maximum number of breaks, whereas the Bai-Perron algorithm only depends on the number of observations. This implies that the Bai-Perron algorithm is better suited for small to moderate samples with a potentially large number of breaks, often found in linear regressions modelling short-run relationships. Instead, the group LARS algorithm is well-suited for large sample sizes and a small to moderate number of structural breaks which is often found for long-run relationships in the presence of structural change.}

\subsection{Asymptotic properties}\label{sec:asymptotics}
\noindent
In the following, we study the asymptotic properties of our adaptive group lasso estimator. To discuss asymptotic properties, we need to impose some further assumptions about the location and magnitude of active breakpoints.

\begin{assumption}\label{as:2}
(i) $I_{\min} = \underset{1 \leq j \leq m_0+1}{\min} \vert t^0_j - t^0_{j-1} \vert > \zeta T$  for some $\zeta > 0$, where $I_{\min}$ is the minimum break interval. \\
(ii) The break magnitudes are bounded to satisfy $m_{\beta} = \underset{1 \leq j \leq m_0 + 1}{\min} \Vert \boldsymbol{\beta}^0_j - \boldsymbol{\beta}^0_{j-1} \Vert > \nu$ for some $\nu > 0$ and $M_{\beta} = \underset{1 \leq j \leq m_0 + 1}{\max} \Vert \boldsymbol{\beta}^0_j - \boldsymbol{\beta}^0_{j-1} \Vert < \infty$. \\
(iii) There exists a constant $C > 0$ such that
\begin{equation*}
\boldsymbol{m}' \boldsymbol{\Sigma} \boldsymbol{m} \geq C \sum\limits_{j \in \bar{\mathcal{A}}}^{} \Vert \boldsymbol{m}_j \Vert^2,
\end{equation*}
for all $TN \times 1$ vectors $\boldsymbol{m} = (\boldsymbol{m}_1, \boldsymbol{m}_2, \dots, \boldsymbol{m}_T)'$ whenever $\sum\limits_{j \in \bar{\mathcal{A}}^c}^{} \Vert \boldsymbol{m}_j \Vert \leq 2 \sum\limits_{j \in \bar{\mathcal{A}}}^{} \Vert \boldsymbol{m}_j \Vert$.
\end{assumption}

Assumption \ref{as:2}(i) requires that the length of the regimes between breaks increases with the sample size and in the same proportions to each other. This allows us to consistently detect and estimate the true break fractions as it makes the break dates asymptotically distinct \citep{Perron2006}. The first inequality of Assumption \ref{as:2}(ii) is a necessary condition to ensure that a structural break occurs at $t_j^0$. We do not consider small breaks with local-to-zero behaviour in this setting (see \cite{BaiLumsdaineStock1998} for assumptions used in this context). This assumption is not believed to be restrictive for the intended empirical applications where applied researchers aim to estimate the long-run equilibrium to obtain the error correction term, i.e., the cointegration residuals, for their follow-up analysis. Essentially, they need optimal in-sample forecasts in terms of mean squared error of the cointegrating regression under structural instability to consistently estimate these residuals. \cite{BootPick2019} show that in-sample forecasts are largely unaffected by local-to-zero breaks. The second part excludes the possibility of infinitely large parameter changes. Assumption \ref{as:2}(iii) implies that the number of active breaks is less than the number of observations and the smallest eigenvalue of $\boldsymbol{\Sigma}_{\bar{\mathcal{A}}}$ is greater than or equal to $C$ by letting $\boldsymbol{m}_j = 0$ for $j \in \bar{\mathcal{A}}^c$. Consequently, Assumption \ref{as:2}(iii) ensures that $\boldsymbol{\Sigma}_{\bar{\mathcal{A}}}$ is positive definite for all $T$. This is only then the case if $\boldsymbol{Z}_{T, \bar{\mathcal{A}}}$ contains columns which are sufficiently distinct. This in turn means that the intervals between breaks need to be sufficiently large for all $T$. It is important to note that Assumption \ref{as:2}(i) can be deduced as an implication of Assumption \ref{as:2}(iii) and we need Assumption \ref{as:2}(iii) exclusively for the first step estimation. Our second step estimation requires only Assumption \ref{as:2}(i) and (ii) as long as consistent weights are available.


First, we need to show that the initial estimator provides consistent weights for the second step adaptive lasso procedure \citep{HuangMaZhang2008}. The following theorem provides a consistency result for the group lasso estimator in cointegrating regressions with (possibly) multiple structural breaks.

\begin{theorem}\label{th:1}
Under Assumption \ref{as:1} and Assumption \ref{as:2}, if $\lambda_T = 2 N c_0 T^{\delta}$ for some $c_0 > 0$ and $3/4 < \delta < 1$, then there exists some $C > 0$ such that with probability greater than $1 - \frac{C}{c_0^2 T^{2\delta - 1}}$,
\begin{equation*}
\Vert \tilde{\boldsymbol{\theta}}(T) - \boldsymbol{\theta}^0(T) \Vert \leq \frac{2}{T^{(1-\delta)/2}} \sqrt{\frac{N c_0 (m_0 + 1) M_{\beta}}{C}}.
\end{equation*}
\end{theorem}

\begin{remark}
The specification of $\lambda_T$ implies that $\lambda_T \to \infty$ for $T \to \infty$. This means we have to apply a stricter penalty for increasing sample sizes to discard a larger set of inactive candidate breaks searching for a fixed number of $m_0$ active breaks. On the other hand, $\lambda_T$ fullfils the condition $\lambda_T/T \to 0$ so that the tuning parameter cannot grow too fast avoiding to ignore active breaks. Since the convergence rate of the group lasso coefficients depends inversely on $\delta$, it is useful to employ a selection rule for $\lambda_T$ where $\delta$ is small. 
\end{remark}

\begin{remark}
Given that $\lambda_T$ is set optimally such that $\delta$ is only slightly above $3/4$, the convergence rate of our first step group lasso estimator is slightly slower than $T^{1/8}$. This means that we lose a substantial portion of the convergence rate which is $T$ for fixed breaks under complete information on their location. The reduced convergence rate can be considered the cost for an estimator which is robust against (multiple) structural breaks with unknown location. For comparison, the convergence rate of in-sample predictions for white noise processes with mean shifts reported in \cite{Harchaoui2010} is $(T / \log T)^{1/4}$. Instead, \cite{ChanYauZhang2014} find that in-sample predictions for piecewise stationary autoregressive processes have a faster convergence rate which amounts to $\sqrt{T / \log T}$, but this result is based on white-noise assumptions for the error term process.
\end{remark}

Theorem \ref{th:1} shows that it is crucial to let the tuning parameter $\lambda_T$ grow at the right rate. However, this rate provides only limited practical guidance towards the choice of $\lambda_T$. We follow \cite{Kock2016}, \cite{QianSu2016} and \cite{SchmidtSchweikert2019} and propose to select $\lambda_T$ by minimizing an information criterion in the form of
\begin{equation}
IC^*(\lambda_{T}) = \log(SSR/T) + \rho_T |\mathcal{A}_T|,
\label{eq:ic}
\end{equation}
where $SSR$ is the sum of squared residuals resulting from the group lasso estimation of Equation~\eqref{eq:gfl} and $|\mathcal{A}_T|$ gives the number of non-zero breakpoint candidates. The penalty function $\rho_T$ allows for different choices. While \cite{Kock2016} suggests to use the BIC for potentially nonstationary autoregressive models which corresponds to $\rho_T = \log(T)/T$, \cite{QianSu2016} propose to use $\rho_T = 1/\sqrt{T}$ for the estimation of structural breaks in stationary time series regressions. In this paper, we follow \cite{SchmidtSchweikert2019} and employ a modified BIC according to \cite{WangLiLeng2009} which incorporates the additional factor $\log \log d^*_T$ where $d^*_T$ denotes the total amount of coefficients in the full model. This modification of the BIC accounts for the fact that the true model must be found in situations where the number of coefficients diverges.

For the next theorem, we temporarily assume that the exact number of breaks is known. This assumption will help us to provide an important consistency result for the estimated location of breakpoints. We note that this temporary assumption will be relaxed for our main results. 
\begin{theorem}\label{th:2}
Under Assumption \ref{as:1} and Assumption \ref{as:2}, if $m_0$ is fixed and $|\mathcal{A}_T| = m_0$, then for all $\epsilon > 0$
\begin{equation*}
P \left( \underset{1 \leq j \leq m_0}{\max} | \hat{t}_j - t^0_j | \leq T \epsilon \right) \to 1, \qquad \text{as } T \to \infty.
\end{equation*}
\end{theorem}

\begin{remark}
Dividing by $T$ on both sides of the inequality in Theorem \ref{th:2} shows that each break fraction can be detected within an $\epsilon$-neighbourhood of its true location. Hence, the convergence rate is similar to the one found in \cite{Davis2006} who use identical assumptions on the minimum break interval. \cite{Harchaoui2010}, allowing for a maximum number of location shifts in white noise processes, report a slightly faster convergence rate. Similarly, \cite{ChanYauZhang2014} apply group lasso to piecewise stationary autoregressive processes with a potentially diverging number of true breakpoints and report the nearly optimal convergence rate $\log T / T$ if errors are Gaussian.
\end{remark}

The previous result is an important building block for our main results. Next, we prove that the group lasso estimator yields a set of estimated breakpoints for which the number of selected breaks is greater than the true number of breaks almost surely when the exact number of breakpoints is unknown. Further, we evaluate the consistency of estimated breakpoints using the Hausdorff distance between the set of estimated breakpoints and the set of true breakpoints. We follow \cite{Boysen2009} and define $d_H(A, B) = \underset{b \in B}{\max} \, \underset{a \in A}{\min} |b - a|$ with $d_H(A, \emptyset) = d_H(\emptyset, B) = 1$, where $\emptyset$ is the empty set. The following theorem shows that the set of estimated breakpoints converges to the set of true breakpoints under the Hausdorff distance.

\begin{theorem}\label{th:3}
If Assumption \ref{as:1} and Assumption \ref{as:2} hold, then as $T \to \infty$
\begin{equation*}
P \left( |\mathcal{A}_T| \geq m_0 \right) \to 1,
\end{equation*}
and for all $\epsilon > 0$
\begin{equation*}
P \left( d_H(\mathcal{A}_T, \mathcal{A}) \leq T\epsilon \right) \to 1.
\end{equation*}
\end{theorem}


\begin{remark}
The first part of Theorem \ref{th:3} yields the familiar result that the group lasso estimator is not model selection consistent in settings where the restricted eigenvalue condition \citep{BickelRitovTsybakov2009} or the irrepresentable condition \citep{Meinshausen2006, ZhaoYu2006} do not hold for the full design matrix. The estimator tends to overselect breakpoints. We note that Assumption \ref{as:2}(iii) is slightly different from the restricted eigenvalue condition used in \cite{BickelRitovTsybakov2009} and restricts only the design submatrix generated from columns containing active breakpoints. This result shows that we do not systematically select too few breaks which is crucial for the intended second step estimation using weights obtained by group lasso estimation. Ignored breaks would directly result in infinite weights for the second step which would mean that these breaks could not be recovered.
\end{remark}

\begin{remark}
The second part of Theorem \ref{th:3} implies that the Hausdorff distance from the set of estimated breakpoints to the true breakpoints diverges slower than the sample size. Consequently, the Hausdorff distance as a percentage of the sample size is bounded by a constant. This provides us with a consistency result for the estimated break fractions and gives us justification to consider multiple structural breaks at once, since the Hausdorff distance evaluates the joint location of all breakpoints.
\end{remark}

Finally, we consider the asymptotic properties of the adaptive group lasso estimator with weights obtained from our first step estimation. We note that Theorem \ref{th:3} allows us to bound the number of breakpoint candidates by a constant. Hence, the dimensionality of the model selection problem no longer depends on the sample size. 

\begin{theorem}\label{th:4}
If Assumption \ref{as:1} and Assumption \ref{as:2} hold, $\lambda_S \to 0$, $\lambda^2_S T^{(1-\delta)\gamma} \to \infty$ for $3/4 < \delta < 1$ and $\gamma > 0$, then \\
(a) Consistency: $\Vert \hat{\boldsymbol{\theta}}_S - \boldsymbol{\theta}^0_S \Vert = O_p(T^{-1})$ \\
(b) Model Selection: $P( g(\lbrace j \geq 2: \Vert \hat{\boldsymbol{\theta}}_{S,j} \Vert \neq 0 \rbrace) = \mathcal{A} ) \to 1$\\
(c) Distribution: 
\begin{equation}
T(\hat{\boldsymbol{\theta}}_{S, \bar{\mathcal{A}}^*} - \boldsymbol{\theta}^0_{S, \bar{\mathcal{A}}^*}) \Rightarrow  \left[ \int\limits_{0}^{1} \boldsymbol{B}_{\tau, \bar{\mathcal{A}}^*}' \boldsymbol{B}_{\tau, \bar{\mathcal{A}}^*} \right]^{-1} \left[ \int\limits_{0}^{1} \boldsymbol{B}_{\tau, \bar{\mathcal{A}}^*} dU + \Lambda^*_{\bar{\mathcal{A}}^*} \right],
\end{equation}
\begin{equation*}
\Lambda^*_{\bar{\mathcal{A}}^*} = \left[\Lambda, (1 - \tau_1) \Lambda, \dots, (1 - \tau_{m_0}) \Lambda \right]', \qquad
\Lambda = \sum\limits_{t=0}^{\infty} E (v_0 u_t),
\end{equation*}
where $\boldsymbol{B}_{\tau, \bar{\mathcal{A}}^*}$ and $U$ are defined in the proof.
\end{theorem}

\begin{remark}
Although the second step tuning parameter $\lambda_S$ can be chosen by a selection rule independent of the first step tuning parameter $\lambda_T$, its value depends on $\delta$, i.e.\ how effective additional coefficients are penalized in the first step and consequently how many truly inactive breakpoint candidates remain in our second step design matrix. Since the number of parameters in the full model can now be limited by a pre-specified maximum number of breaks, we suggest to use an information criterion like the BIC, which has performed quite well in our simulation experiments. 
\end{remark}

\begin{remark}
Combining parts (a) to (c) of  Theorem \ref{th:4} shows that the adaptive group lasso estimator has oracle properties. This means that the adaptive group lasso performs correct model selection and has the same asymptotic distribution as the least squares estimator if the breaks' location would have been known beforehand. Since our regression involves nonstationary components, the asymptotic distribution of the least squares estimator is naturally given as a functional of Brownian motions. \cite{SchmidtSchweikert2019} use the term `nonstandard oracle property' to distinguish it from the term used in \cite{FanLi2001}. The asymptotic bias term $\Lambda$ originating from the dependency between increments of the regressors and the error term of the cointegrating regression can be eliminated using dynamic augmentation according to \cite{Saikkonen1991} and \cite{StockWatson1993}. 
\end{remark}

\begin{remark}
It is notable that our estimator has nonstandard oracle properties although the convergence rate of the group lasso estimator is slower than $T^{1/8}$. \cite{Zou2006} argues that the convergence rate of the initial estimator is allowed to be substantially slower than the desired convergence rate of the adaptive lasso estimator if the tuning parameter is specified accordingly.
\end{remark}

\section{Summary of Monte Carlo Experiments}\label{sec:sim}
\noindent
In this section, we conduct simulation experiments to assess the adequacy of our technical results in \autoref{sec:method}. We investigate the finite sample performance of our adaptive group lasso procedure with respect to the accuracy in finding the exact number of breaks, their location and the magnitude of parameter changes. We consider model specifications with one, two and four breakpoints, respectively. The following DGP is employed to model a multivariate cointegrated system with multiple structural breaks, 
\begin{equation}
\label{eq:mc.dgp}
\begin{array}{lllllll}
y_t & = & \mu + \boldsymbol{\beta}_t \boldsymbol{X}_t + \vartheta_t & & \vartheta_t & \sim & N(0,\sigma^2_{\vartheta}), \\
\boldsymbol{X}_t & = &  \boldsymbol{X}_{t-1} + \omega_t & & \omega_t & \sim & N(0,\Sigma), \\
\end{array}
\end{equation}
where $\boldsymbol{X}_t = (X_{1t}, X_{2t}, \dots, X_{Nt})'$ and $\Sigma = diag(\sigma^2_{\omega})$, i.e.\ the innovations of our generated random walk processes have identical normal distributions. $\mu$ is a non-zero intercept and $\boldsymbol{\beta}_t = (\beta_{1t}, \beta_{2t}, \dots, \beta_{Nt})$ is a time-varying slope coefficient vector with non-zero baseline value and a finite number of breaks. We note that $cov(\vartheta_t, \omega_t) = 0$, i.e.\ our regressors are strictly exogenous and the asymptotic bias reported in Theorem \ref{th:4} is non-existent.

Naturally, the ability of all structural break estimators to detect breaks depends on the overall signal strength. \cite{NiuHaoZhang2015} define signal strength in change-point models by $S = m_{\beta}^2 I_{\min}$, where $I_{\min} = \underset{1 \leq j \leq m_0+1}{\min} \vert t_j - t_{j-1} \vert$ is the minimum distance between breaks and $m_{\beta} = \underset{1 \leq j \leq m_0 + 1}{\min} \Vert \boldsymbol{\beta}_j - \boldsymbol{\beta}_{j-1} \Vert$ is the minimum jump size. For our main simulations concerned with consistency of the adaptive group lasso estimator, we use equal jump sizes for multiple breaks and locate the breaks with equidistant spacing between them. Hence, overall signal strength is a linear function of the sample size in our simulations. We use a baseline value of two and a jump size of two which is equal to the standard deviation of the regression error term. Simulations with a better signal-to-noise ratio yield more precise estimates for all sample sizes. 

In \autoref{tab:sb_diverg_sig2}, we report our results for $N = 2$ regressors. We specify our model for one break located at $\tau = 0.5$, two breaks at $\tau = (0.33, 0.67)$ and four breaks at $\tau = (0.2, 0.4, 0.6, 0.8)$ to have an equidistant spacing on the unit interval. We first compute the percentages of correct estimation (pce) of the number of breaks $m$ and measure the accuracy of the break date estimation conditional on the correct estimation of $m$. For this matter, we compute the average Hausdorff distance and divide it by $T$ (hd/T) to compare the values across different sample sizes. The corresponding figures in our tables are reported in percentages. As $T$ grows larger, the number of breaks is detected with increasing precision and the distance between estimated breakpoints and true breakpoints declines to nearly zero. Parameter estimates are already very accurate at small sample sizes. As expected, the parameter changes of models with fewer breakpoints can be estimated more precisely than those of models with a larger number of breakpoints, as indicated by larger standard deviations obtained for the latter at all sample sizes.\footnote{Results for $N = 1$ reported in \cite{SchmidtSchweikert2019} show a very similar pattern. We find that it is slightly more difficult to detect the correct number of breaks in regressions with multiple regressors although the jump size measured as the Euclidean distance is equal for both settings.} Comparing these results with those obtained for the Bai-Perron algorithm\footnote{\cite{KejriwalPerron2008, KejriwalPerron2010} obtain estimates of the parameters using the dynamic programming algorithm of \cite{BaiPerron2003} with no modification since the algorithm itself is valid irrespective of the nature of the regressors and errors given that it detects break dates that minimize the global sum of squared residuals in a regression.}, where the number of breaks is determined via the BIC, we find that both approaches perform similarly well. The results are reported in \autoref{tab:sb_diverg_sig2_baiperron}. While the Bai-Perron algorithm estimates the true break fractions slightly more accurately, parameter changes on average have larger standard deviations at all samples sizes. The number of structural breaks is estimated with identical accuracy.\footnote{ We can confirm the theoretical claims made about the computational complexity of the Bai-Perron algorithm in comparisons to the group LARS algorithm in \autoref{sec:framework}. We obtain the following computational times (in seconds) for both algorithms. First, using the simulation set-up for \autoref{tab:sb_diverg_sig2} and a sample size of $T = 1000$, we have $M = 1$: (gLARS: 2.61, BP: 11.03), $M = 2$: (gLARS: 9.24, BP: 11.41), $M = 4$: (gLARS: 21.27, BP: 15.36). Here, we find that the Bai-Perron algorithm is more robust to a larger number of breaks in terms of computational time. Second, we increase the sample size to $T = 10000$ and record the following times, $M = 1$: (gLARS: 4.04, BP: 1013.15), $M = 2$: (gLARS: 27.35, BP: 1317.45), $M = 4$: (gLARS: 35.23, BP: 1563.55). In this case, we can confirm that the group LARS algorithm is much more computationally efficient for large sample sizes. All simulations are computed on a computer with an Intel i5-6500 CPU at 3.20GHz and 16GB RAM.}


Next, we investigate if dynamic augmentation according to \cite{Saikkonen1991} and \cite{StockWatson1993} yields consistent coefficient estimates if the strict exogeneity condition of our main results is violated. To do so, we follow \cite{KejriwalPerron2008} and draw the vector $(\vartheta_t, \omega_{1t}, \omega_{2t})'$ jointly from a multivariate normal distribution with zero mean and covariance matrix
\begin{equation}
V = \begin{pmatrix}
\sigma^2_{\vartheta} & 0.5 & 0.5 \\
0.5 & \sigma^2_{\omega} & 0 \\
0.5 & 0 & \sigma^2_{\omega}
\label{eq:V}
\end{pmatrix}.
\end{equation}
Using this configuration, the strict exogeneity condition is violated for both regressors but the regressors are still generated by independent processes. If we attempt to detect and estimate structural breaks without dynamic augmentation, we still detect breakpoints precisely but obtain strongly biased coefficient estimates. In \autoref{tab:sb_endogenous_sig2}, we find the corresponding results after the inclusion of $l=1$ and $l=2$ leads and lags. Now, we can recover the number, location and magnitude of all breakpoints with similar accuracy compared to our simulations under strict exogeneity.

In \autoref{tab:sb_partial_sig2}, we consider partial breaks in the cointegrating vector. We use a model specification according to the DGP in Equation~\eqref{eq:mc.dgp} with $N = 2$ regressors and induce partial structural breaks through $\beta_{1t}$ only. Our estimator is applied estimating a full structural change model without prior knowledge that $\beta_{2t}$ is constant over the sampling period. Again, we observe that the number of breaks, their timing and their magnitude is consistently estimated.\footnote{Comparing the results with those obtained for the Bai-Perron algorithm (not reported), we again find that both approaches detect the number of structural breaks with identical accuracy. The Bai-Perron algorithm estimates the true break fractions slightly more accurately, but parameter changes have larger standard deviations at for samples sizes $T=200$ and $T=400$.} The distance between the set of estimated breakpoints and true breakpoints is larger than in the full break setting in \autoref{tab:sb_diverg_sig2}. This result is not surprising considering that the break magnitudes for partial breaks are smaller making it more difficult for the adaptive group lasso procedure to detect the true location of the breaks. Consequently, these results also help us to assess how the break magnitude influences the detection rates. Reducing the Euclidean distance from 2 to $\sqrt{2}$, roughly doubles the average Hausdorff distance. The convergence rates for zero parameter changes in $\beta_{2t}$ is almost identical to the convergence rate observed for the non-zero parameter changes in $\beta_{1t}$. This is naturally driven by the joint evaluation of all regressors in each group. Unlike bi-level estimators proposed in \cite{HuangMaXieZhang2009} and \cite{BrehenyHuang2009}, the adaptive lasso procedure is not able to shrink coefficients within active groups to zero. Hence, the usual convergence rate for non-zero coefficients apply. In these cases, the convergence rate for $\beta_{2t}$ could in principle be increased if our procedure was extended to feature bi-level shrinkage. However, this is beyond the scope of this paper and is not investigated further at this point.

Finally, we investigate how sensitive our procedure is to break fractions located near the boundary of the unit interval. While the properties of tests for structural changes in the literature depend strongly on the trimming parameter \citep{BaiPerron2006}, our method to recover breaks should be more robust in this regard. We only need some lateral trimming to ensure that the first and last regimes identified by our adaptive lasso procedure comprise a sufficiently large number of observations to estimate regime-dependent coefficients.\footnote{For our main results, reported in \autoref{tab:sb_diverg_sig2} and \autoref{tab:sb_diverg_sig2_baiperron}, we follow \cite{KejriwalPerron2008} and use a 15\% lateral trimming to compare both methods.} The results for breaks near the boundary are summarized in \autoref{tab:sb_bounds_sig2}. The first and second panel considers one break located at $\tau = 0.1$ and $\tau = 0.9$, respectively. The pce and average Hausdorff distances over all sample sizes clearly show that a break located close to the beginning of the sample is more difficult to detect than a break located at the end of the sample. \cite{GregoryHansen1996} and \cite{Schweikert2019} report similar findings for their grid search algorithms. To investigate this further, we consider two breaks located at $\tau = (0.1, 0.9)$ in panel three of \autoref{tab:sb_bounds_sig2}. Here, we find that the pce is quite low compared to our main results with equidistant spacing of breakpoints. The first break is estimated less accurately than the second break which can be explained by the fact that parameter changes are measured from one regime to the next and that only a relatively small number of observations is available to estimate the break at $\tau = 0.1$. 

The results of our first series of boundary experiments imply that it might be possible to relax our trimming restrictions and assume an asymmetric lateral trimming where the first regime must contain sufficiently many observation, say 5\% of the sample, while the end of the sample does not necessarily have to be excluded. We apply a 0.05/0 trimming and estimate breaks located at $\tau = (0.1, 0.95)$. The results for this trimming strategy are presented in panel four of \autoref{tab:sb_bounds_sig2}. The break at $\tau = 0.95$ can still be accurately detected, however the standard errors of the parameter changes increase due to the smaller number of observations in the last regime. We conclude that trimming is not necessary to detect breaks located at the end of the sample. Still, we suggest to set a minimum number of observations per regime to ensure that parameter changes are estimated precisely.

\section{Application: US Money Demand}\label{sec:app}
\noindent
In this section, we apply our proposed methodology to the US money demand function. Particularly, we estimate a long-run money demand specification and investigate the presence of long-run instabilities in a cointegrating framework. \cite{Juselius2006} considers the condition $M/P = L(Y, R)$ for equilibrium in the money market, which relates $M/P$, the ratio of nominal money balances to price levels, to real income $Y$ and the short term nominal interest rate $R$. Two competing empirical specifications are considered in the literature, namely, the semi-log and the log-log specification. The latter is given by $L(Y, R) = \alpha Y^{\beta_1} R^{\beta_2}$, where $\alpha$ is a constant, $\beta_1$ is the income-elasticity assumed to be unity and $\beta_2 < 0$ is the interest-elasticity.\footnote{Recall that in general, coefficients of log-transformed variables in cointegrating regressions should not be interpreted as elasticities \citep{Johansen2005}. Only in the special case when those variables are strongly exogeneous, it is allowed to use a ceteris paribus interpretation for the corresponding coefficients.} For our empirical application, we choose a log-log specification which has been found to fit quite well to US data \citep{Lucas2000, Bae2007, Ireland2009, MoglianiUrga2018}. We extend the dataset used by \cite{Maki2012} to span the period from January 1959 to December 2018. Monthly data are obtained from the Federal Reserve Bank of St. Louis. We consider the empirical US money demand function,
\begin{equation}
m^*_t = \mu + \beta_1 y_t + \beta_2 r_t + u_t,
\label{eq:money}
\end{equation}
where $m^*_t$ and $y_t$ denote the natural logarithm of the ratio of nominal money balances to price levels, and the natural logarithm of real income, respectively. According to the log-log specification, we employ the natural logarithm of the short term nominal interest rate, denoted by $r_t$. $u_t$ denotes the equilibrium error of the money demand function if the system is cointegrated. We use M2 as nominal money, the consumer price index as prices, and the index of industrial production as real income. For the interest rate, we use the 6-month Treasury bill rate. All time series are tested for a unit root using the Dickey-Fuller test. The results, which are not reported, support the assumption that all variables are integrated of order one and we can continue our cointegration analysis.

First, we assume constancy of the parameters and ignore potential structural breaks. Estimation of the long-run equilibrium equation yields coefficients $\hat{\mu} = -0.05$, $\hat{\beta}_1 = 0.80$ and $\hat{\beta}_2 = -0.08$. Dynamic augmentation of the cointegrating regression with two leads and lags each, does not change the coefficient values. The Engle-Granger test based on an ADF regression yields the t-ratio $-0.063$ which does not lead to a rejection of the null hypothesis at the 10\% level. Similar results can be obtained for the Phillips-Ouliaris test and the Johansen test. Although it is implausible from a theoretical standpoint that the system is not cointegrated, at least our estimated coefficients have the expected sign and magnitude for post-war data. The estimated income-elasticity measured by $\beta_1$ is slightly below the theoretically expected value. The interest-elasticity of money demand, measured by $\beta_2$ is expected to be negative. \cite{Lucas2000} considers $-0.3$, $-0.5$ and $-0.7$ as values of $\beta_2$ and finds that $\beta_2 = -0.5$ gives the best fit for US data. \cite{Meltzer1963}, \cite{Lucas1988}, \cite{HoffmanRasche1991} and \cite{StockWatson1993} find empirical evidence consistent with the theoretical expectation that income-elasticity of money demand is unity and interest-elasticity is relatively high. \cite{Ball2001} studies subperiods from 1903 to 1994 and argues against a stable long-run money demand. Further empirical studies have pointed out the presence of structural instability in US money demand for sample periods including data from the 1990s and 2000s \citep{TelesZhou2005, Wang2011, LucasNicolini2015}. Potential nonlinearities in the functional form are investigated, for example, by \cite{ChenWu2005} and \cite{JawadiSousa2013}. However, we take the perspective that the linear cointegrating regression in Equation~\eqref{eq:money} approximates the data well if we simultaneously account for (multiple) parameter changes during the sample period.

\begin{figure}[h!]
\centering
\includegraphics[scale=0.65]{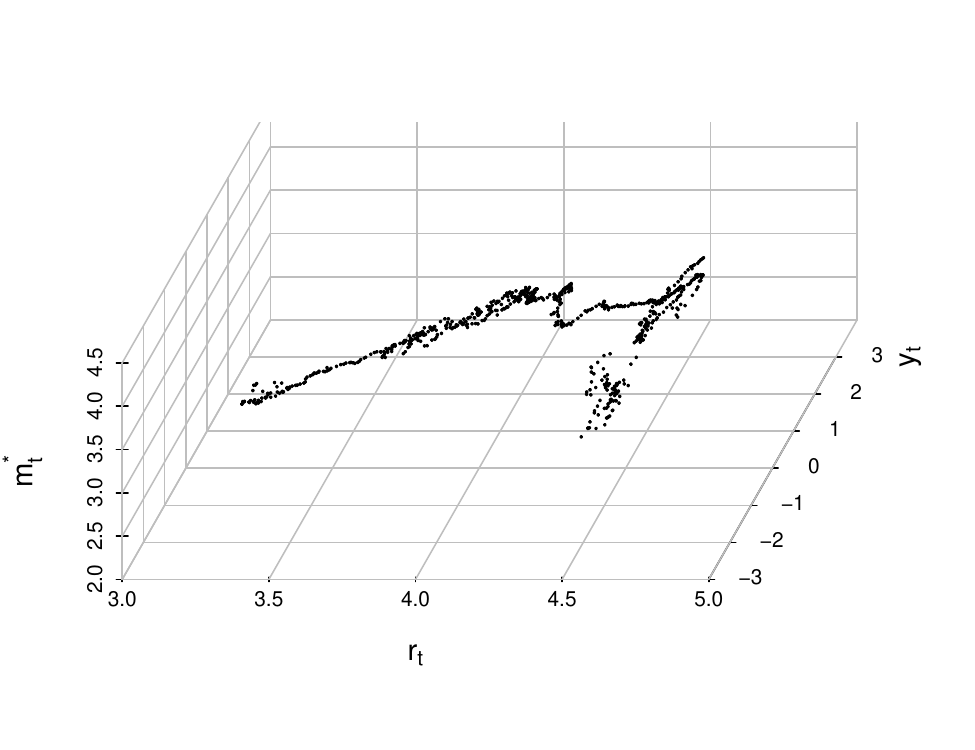}
\captionof{figure}{\small{Three-dimensional scatterplot of $r_t$ (x-axis), $y_t$ (y-axis) and $m^*_t$ (z-axis).}}
\label{fig:scatter3d}
\end{figure}
\begin{figure}[h!]
\centering
\includegraphics[scale=0.65]{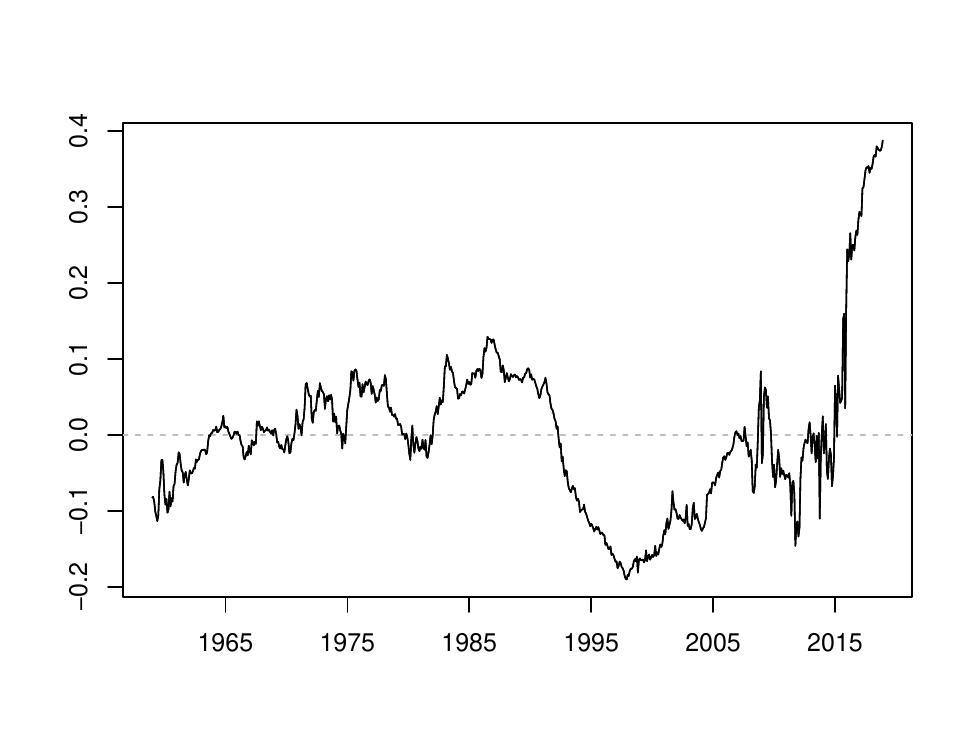}
\captionof{figure}{\small{Residual series obtained from least squares estimation.}}
\label{fig:resid}

\centering
\includegraphics[scale=0.65]{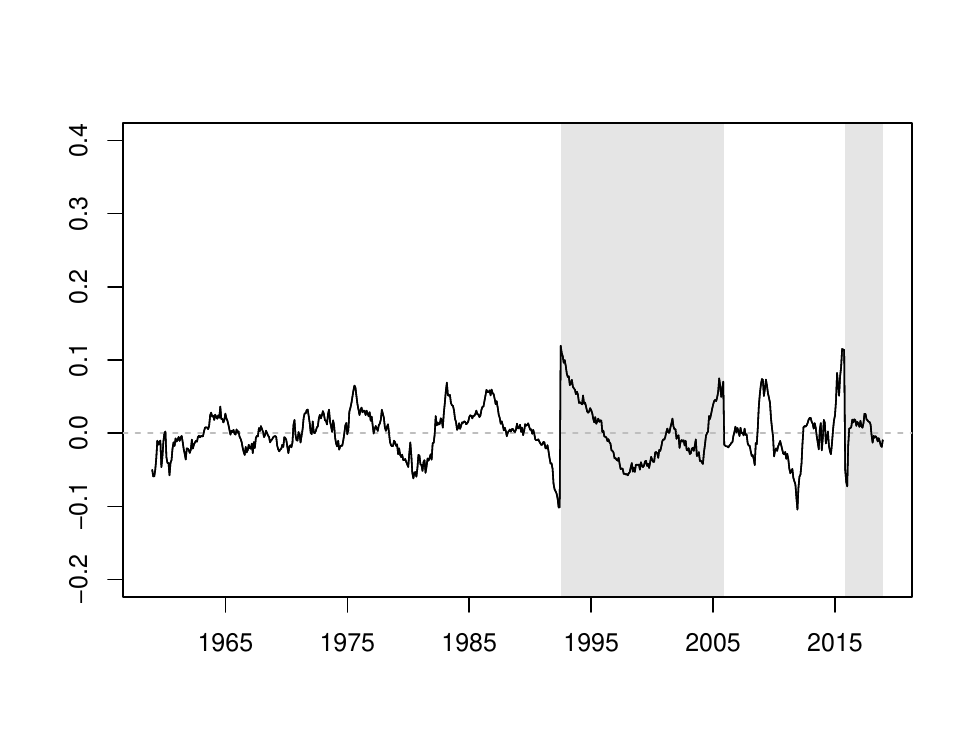}
\captionof{figure}{\small{Post-lasso residual series. Estimated regimes are marked by grey and white areas.}}
\label{fig:lasso_resid}
\end{figure}

A three-dimensional scatterplot of the data in \autoref{fig:scatter3d} reveals that the relationship between $r_t$,  $y_t$ and $m^*_t$ has changed during the sampling period. We observe at least three two-dimensional surfaces which correspond to distinct long-run levels from which $m^*_t$ does not persistently deviate. However, if we consider linear cointegration without the possibility of structural breaks, we infer from \autoref{fig:resid} that the residual series exhibits a clear trend during the latter half of the sample. We note that the presence of structural breaks might mask the cointegrating relationship. Next, we compare several previously mentioned structural break models with our model selection approach. The \cite{GregoryHansen1996} test indicates a breakpoint at 2008 m06 but does not reject the null hypothesis at the 10\% level. Because the GH-test does not model structural breaks under the null hypothesis, this means that the timing of the indicated breakpoint is not informative. The \cite{Hatemi-J2008} test indicates two breakpoints at 1992 m01 and 2008 m06. The null hypothesis of no cointegration can be rejected at the 5\% level if these breakpoints are taken into account. The maximum number of breaks chosen for the \cite{Maki2012} test is five. It selects the breakpoints at 1986 m05, 1992 m04, 2004 m05, 2008 m11, 2014 m03 and rejects the null hypothesis of no cointegration at the 1\% level. We initially also start with a maximum of five breakpoints for our adaptive group lasso procedure. However, imposing a minimum regime length of one year to precisely estimate the parameter changes and dynamically augmenting the cointegrating regression results in a model specification with three breakpoints. The final estimates yield break dates 1992 m07, 2005 m12, and 2015 m11.\footnote{The corresponding breakpoint estimates using the Bai-Perron algorithm are almost identically located at 1991 m10, 2004 m07, and 2014 m06.} 

The income-elasticity from 1959 m01 to 1992 m07 is estimated to be $0.95$ and the interest-elasticity amounts to $-0.10$ for the same period. These estimates correspond to the theoretical predictions formulated in \cite{Juselius2006} and to the results reported in empirical papers considering this sample period \citep{Lucas1988, StockWatson1993, Lucas2000}. The first breakpoint leads to an income-elasticity reduction from $0.95$ to $0.89$ while the interest-elasticity remains largely unchanged. A partial decoupling of money demand from income might be explained by the begin of the costly Gulf War and a sharp increase in US debt. In turn, the second breakpoint at 2005 m12 has a negligible effect on the income-elasticity ($0.89$ to $0.90$) but results in a larger reduction of the interest-elasticity from $-0.10$ to $-0.07$. This breakpoint can be related to the beginning Global Financial Crisis of 2007-2008. It must be emphasized at this point that estimated break dates might be affected by the usual lead and lag effects, since parameter changes are representative for the following regime. In the aftermath of the Global Financial Crisis, the Federal Reserve implemented a zero interest rate policy. Consequently, the variation in the interest rate for this period approached zero which naturally reduced the interest-elasticity of money demand. After 2015 m11, the expected interest-elasticity does no longer achieve a good fit to the data and increases to $0.01$. In contrast, the income-elasticity is very close to unity ($0.97$).

Accounting for structural breaks, as indicated by the adaptive group lasso procedure, yields a residual series which much more resembles being generated by a stationary process than the original OLS residual. \autoref{fig:lasso_resid} illustrates that the residual series does not exhibit a visible trend. The speed of adjustment after equilibrium errors is now $-0.097$ which means that roughly 10\% of long-run deviations are corrected each period.

\section{Conclusion}\label{sec:conc}
\noindent
In this paper, we propose a penalized regression approach to the problem of detecting an unknown number of structural breaks and their location in cointegrating regressions. Our estimator eliminates irrelevant breakpoints from a set of candidate breakpoints and, hence, follows a top-down approach regarding the estimation of structural breaks. Practitioners should apply this new methodology in complement to the Bai-Perron algorithm which follows a bottom-up approach, i.e.\ sequentially increasing the number of breaks. Due to the importance of finding the right model specification with respect to the number and location of structural breaks, either approach can serve as a valuable robustness check of the model specification chosen by the other approach. Ideally both approaches should indicate the same breakpoints which would mean that the chosen model specification is sufficiently sparse (bottom-up) and does not ignore important breaks (top-down).

We can show the important theoretical result that the adaptive group lasso estimator has nonstandard oracle properties in settings with a diverging number of breakpoint candidates. This means that the estimator determines the true number of non-zero parameter changes with probability tending to one and consistently estimates their location. The corresponding parameter changes are estimated with the same convergence rate that least squares estimators would have under full information of the number and location of breaks. 

The present paper does not consider cointegration testing. It is unclear how optimal cointegration test can be constructed from the proposed penalized regression approach. An attempt to design such cointegration tests has been made by \cite{SchmidtSchweikert2019} for a single regressor. Our results depend critically on the stationarity assumption about the error term. Hence, it is required to establish the existence of a cointegration relationship before the penalized regression is estimated. Practitioners should employ cointegration tests which are robust to the presumed number of breaks during the sample period.

Further extensions include the use of bi-level selection via the group fused lasso \citep{HuangMaXieZhang2009, BrehenyHuang2009} to estimate partial breaks more efficiently, and the possibility to detect structural breaks in system-based approaches with multiple equilibria \citep{BaiLumsdaineStock1998, Qu2007}.

\section{Acknowledgements}
\noindent
I thank Florian Stark, Alexander Schmidt, Markus Mößler, Timo Dimitriadis and the participants of the Doctoral Seminar in Econometrics in Tübingen, German Statistical Week in Trier, ZU Methodenkolloquium in Friedrichshafen, THE Christmas Workshop in Stuttgart, Seminar at Maastricht University, and the 2nd CSL Symposium in Stuttgart for valuable comments and suggestions. Further, I thank Maike Becker and Manuel Huth for excellent research assistance.

\clearpage

\appendix

\section{Mathematical Appendix}\label{sec:appendix}

\renewcommand{\theequation}{\thesection.\arabic{equation}}
\setcounter{equation}{0}

\begin{lemma}\label{lemma:1}
Under Assumption \ref{as:1} and Assumption \ref{as:2}, for any $c_0 > 0$ and $\delta > 1/2$, there exists some constant $C > 0$ such that
\begin{equation}
P \left( \underset{1 \leq s \leq T}{\max} \, \underset{1 \leq j \leq N}{\max} \left\vert \frac{1}{T} \sum\limits_{i=s}^{T} X_{ji} u_i \right\vert \geq c_0 T^{\delta} \right) \leq \frac{C}{c_0^2 T^{2\delta - 1}}.
\label{eq:lemma1}
\end{equation}
\end{lemma}

\begin{proofoflemma}
According to Theorem 4.1 of \cite{Hansen1992a}, it holds for all $j = 1, \dots, N$ and $0 \leq r \leq 1$ that $T^{-1} \sum\limits_{i=1}^{\lbrack Tr \rbrack} X_{ji} u_i$ has weak limit $\int\limits_{0}^{r} B_j dU + r\Lambda_j$, where $X_{j[Tr]} = T^{-1/2} \sum\limits_{t=1}^{\lbrack Tr \rbrack} v_{jt} \Rightarrow B_j(r)$, $\Lambda_j = \sum\limits_{j=0}^{\infty} E v_0 u_j$ and $B_j(r)$ is a Brownian motion process with variance $\sigma^2$. Since the second moment of $\int\limits_{0}^{r} B_j dU + r\Lambda_j$ is finite and $T^{-2} \left( \sum\limits_{i=1}^{\lbrack Tr \rbrack} X_{ji} u_i \right)^2$ is uniformly integrable, we have $E \left( \left\vert \frac{1}{T} \sum\limits_{i=1}^{s} X_{ji} u_i \right\vert^2 \right) \leq \frac{C}{8}$ for $1 \leq s \leq T$ and all $T$ according to Theorem 3.5 of \cite{Billingsley1999}. It follows from Markov inequality that
\begin{eqnarray}
\notag P \left( \underset{1 \leq s \leq T}{\max} \left\vert \frac{1}{T} \sum\limits_{i=1}^{s} X_{ji} u_i \right\vert \geq \frac{c_0 T^{\delta}}{2} \right) &\leq& \sum\limits_{s=1}^{T} P \left( \left\vert  \frac{1}{T} \sum\limits_{i=1}^{s} X_{ji} u_i \right\vert \geq \frac{c_0 T^{\delta}}{2} \right) \\
 &\leq& \sum\limits_{s=1}^{T} \frac{4}{c_0^2 T^{2\delta}} E\left( \left\vert \frac{1}{T} \sum\limits_{i=1}^{s} X_{ji} u_i \right\vert^2 \right)\\
\notag &\leq& \frac{C}{2c_0^2 T^{2\delta-1}},
\end{eqnarray}
for some $C > 0$. Thus, it holds that
\begin{eqnarray}
\notag && P \left( \underset{1 \leq s \leq T}{\max} \left\vert \frac{1}{T} \sum\limits_{i=s}^{T} X_{ji} u_i \right\vert \geq c_0 T^{\delta} \right) \\
\notag &=& P \left( \underset{1 \leq s \leq T}{\max} \left\vert \frac{1}{T} \sum\limits_{i=1}^{T} X_{ji} u_i - \frac{1}{T} \sum\limits_{i=1}^{s-1} X_{ji} u_i \right\vert \geq c_0 T^{\delta} \right) \\
&\leq& P \left( \left\vert \frac{1}{T} \sum\limits_{i=1}^{T} X_{ji} u_i \right\vert \geq \frac{c_0}{2} T^{\delta} \right) + P \left( \underset{1 \leq s \leq T}{\max} \left\vert \frac{1}{T} \sum\limits_{i=1}^{s-1} X_{ji} u_i \right\vert \geq \frac{c_0}{2} T^{\delta} \right) \\
\notag &\leq& \frac{C}{c_0^2 T^{2\delta - 1}}.
\end{eqnarray}
Since $N$ is finite, Equation~\eqref{eq:lemma1} follows.
\end{proofoflemma}
$\hfill \Box$


\begin{lemma}\label{lemma:2}
Let $\tilde{\boldsymbol{\theta}}(T)$ be the estimator of $\boldsymbol{\theta}(T)$ as defined in Equation~\eqref{eq:gfl}, then it holds under the same conditions as in Theorem \ref{th:1} that
\begin{equation*}
\sum\limits_{s = \hat{t}_j}^{T} \boldsymbol{X}_s \left( y_s - \sum\limits_{i=1}^{s} \tilde{\boldsymbol{\theta}}_i' \boldsymbol{X}_s \right) - \frac{1}{2} T \lambda_T \frac{\tilde{\boldsymbol{\theta}}_{\hat{t}_j}}{\Vert \tilde{\boldsymbol{\theta}}_{\hat{t}_j} \Vert} = 0, \qquad \forall \tilde{\boldsymbol{\theta}}_{\hat{t}_j} \neq 0,
\end{equation*}
and 
\begin{equation*}
\Vert \sum\limits_{s = j}^{T} \boldsymbol{X}_s \left( y_s - \sum\limits_{i=1}^{s} \tilde{\boldsymbol{\theta}}_i' \boldsymbol{X}_s \right) \Vert \leq \frac{1}{2} T \lambda_T, \qquad \forall j. 
\end{equation*}
\end{lemma}
Furthermore $\sum\limits_{i=1}^{t} \tilde{\boldsymbol{\theta}}_i = \tilde{\boldsymbol{\beta}}_j$ for $\hat{t}_{j-1} \leq t \leq \hat{t}_j - 1$, $j = 1, 2, \dots, |\mathcal{A}_T|$.

\begin{proofoflemma}
This lemma is a direct consequence of the Karush-Kuhn-Tucker (KKT) conditions for group lasso estimators.
\end{proofoflemma}
$\hfill \Box$

\begin{lemma}\label{lemma:3}
A necessary and sufficient condition for the estimator $\hat{\boldsymbol{\theta}}_S$ to be a solution to the adaptive group lasso objective function $Q(\boldsymbol{\theta})$ is
\begin{equation*}
\boldsymbol{Z}_{S,i} \left( y_T - \boldsymbol{Z}_S \hat{\boldsymbol{\theta}}_S \right) - \frac{1}{2} T \lambda_S \Vert \tilde{\boldsymbol{\theta}}_{S,i} \Vert^{-\gamma} \frac{\hat{\boldsymbol{\theta}}_{S,i}}{\Vert \hat{\boldsymbol{\theta}}_{S,i} \Vert} = 0, \qquad \forall \hat{\boldsymbol{\theta}}_{S,i} \neq 0,
\end{equation*}
and 
\begin{equation*}
\Vert \boldsymbol{Z}_{S,i} \left( y_T - \boldsymbol{Z}_S \hat{\boldsymbol{\theta}}_S \right) \Vert \leq \frac{1}{2} T \lambda_S \Vert \tilde{\boldsymbol{\theta}}_{S,i} \Vert^{-\gamma}, \qquad \forall \hat{\boldsymbol{\theta}}_{S,i} = 0.
\end{equation*}
\end{lemma}

\begin{proofoflemma}
This lemma is a direct consequence of the Karush-Kuhn-Tucker (KKT) conditions for adaptive group lasso estimators.
\end{proofoflemma}
$\hfill \Box$


\begin{proofof}
We prove that the group lasso estimator consistently estimates all coefficients. The first part of our proof is related to the results given in \cite{ChanYauZhang2014}, while the second part uses ideas presented in \cite{HeHuang2016}. By definition of $\tilde{\boldsymbol{\theta}}(T)$, it holds that
\begin{equation}
\frac{1}{T} \Vert y_T - \boldsymbol{Z}_T \tilde{\boldsymbol{\theta}}(T) \Vert^2 + \lambda_T \sum\limits_{i=1}^{T} \Vert \tilde{\boldsymbol{\theta}}_i \Vert \leq \frac{1}{T} \Vert y_T - \boldsymbol{Z}_T \boldsymbol{\theta}^0(T) \Vert^2 + \lambda_T \sum\limits_{i=1}^{T} \Vert \boldsymbol{\theta}^0_i \Vert.
\label{eq:p.argmin}
\end{equation}
Note that $\bar{\mathcal{A}}$ contains the indices of all truly non-zero coefficients. Inserting $y_T = \boldsymbol{Z}_T \boldsymbol{\theta}^0(T) + u_T$ into Equation~\eqref{eq:p.argmin} yields
\begin{eqnarray}
\notag &&\frac{1}{T} \Vert \boldsymbol{Z}_T (\boldsymbol{\theta}^0(T) - \tilde{\boldsymbol{\theta}}(T)) \Vert^2 \\
&\leq& 2 \sum\limits_{j=1}^{T} (\tilde{\boldsymbol{\theta}}_j - \boldsymbol{\theta}^0_j)' \left( \frac{1}{T} \sum\limits_{i=j}^{T} \boldsymbol{X}_i u_i \right) + \lambda_T \sum\limits_{i \in \bar{\mathcal{A}}}^{} (\Vert \boldsymbol{\theta}^0_i \Vert - \Vert \tilde{\boldsymbol{\theta}}_i \Vert) - \lambda_T \sum\limits_{i \in \bar{\mathcal{A}}^c}^{} \Vert \tilde{\boldsymbol{\theta}}_i \Vert \\
\notag &\leq& 2 N \left( \sum\limits_{j=1}^{T} \Vert \tilde{\boldsymbol{\theta}}_j - \boldsymbol{\theta}^0_j \Vert \right) \left( \underset{1 \leq l \leq N}{\max} \left\vert \frac{1}{T} \sum\limits_{i=j}^{T} \boldsymbol{X}_{li} u_i \right\vert \right) + \lambda_T \sum\limits_{i \in \bar{\mathcal{A}}}^{} (\Vert \boldsymbol{\theta}^0_i \Vert - \Vert \tilde{\boldsymbol{\theta}}_i \Vert) - \lambda_T \sum\limits_{i \in \bar{\mathcal{A}}^c}^{} \Vert \tilde{\boldsymbol{\theta}}_i \Vert.
\end{eqnarray}
Noting that
\begin{equation}
\sum\limits_{j=1}^{T} \Vert \tilde{\boldsymbol{\theta}}_j - \boldsymbol{\theta}^0_j \Vert = \sum\limits_{j \in \bar{\mathcal{A}}}^{} \Vert \tilde{\boldsymbol{\theta}}_j - \boldsymbol{\theta}^0_j \Vert +  \sum\limits_{j \in \bar{\mathcal{A}}^c}^{} \Vert \tilde{\boldsymbol{\theta}}_j \Vert,
\label{eq:theta}
\end{equation}
and using Lemma \ref{lemma:1}, we have with probability greater than $1 - \frac{C}{c_0^2 T^{2\delta - 1}}$ that
\begin{eqnarray}
\notag &&\frac{1}{T} \Vert \boldsymbol{Z}_T (\boldsymbol{\theta}^0(T) - \tilde{\boldsymbol{\theta}}(T)) \Vert^2 \\
\notag &\leq& 2 N c_0 T^{\delta} \left( \sum\limits_{j=1}^{T} \Vert \tilde{\boldsymbol{\theta}}_j - \boldsymbol{\theta}^0_j \Vert \right) + \lambda_T \sum\limits_{i \in \bar{\mathcal{A}}}^{} (\Vert \boldsymbol{\theta}^0_i \Vert - \Vert \tilde{\boldsymbol{\theta}}_i \Vert) - \lambda_T \sum\limits_{i \in \bar{\mathcal{A}}^c}^{} \Vert \tilde{\boldsymbol{\theta}}_i \Vert \\
&=& \lambda_T \sum\limits_{i \in \bar{\mathcal{A}}}^{} \Vert \tilde{\boldsymbol{\theta}}_i - \boldsymbol{\theta}^0_i \Vert + \lambda_T \sum\limits_{i \in \bar{\mathcal{A}}}^{} \left( \Vert \boldsymbol{\theta}^0_i \Vert - \Vert \tilde{\boldsymbol{\theta}}_i \Vert \right) \\
\notag &\leq& 2 \lambda_T \sum\limits_{i \in \bar{\mathcal{A}}}^{} \Vert \boldsymbol{\theta}^0_i \Vert \leq 2 \lambda_T (m_0 + 1) \underset{1 \leq j \leq m_0+1}{\max} \, \Vert \boldsymbol{\beta}^0_j - \boldsymbol{\beta}^0_{j-1} \Vert = 4 N c_0 T^{\delta} (m_0 + 1) M_{\beta},
\label{eq:prederr}
\end{eqnarray}
where both equalities follow from the definition of $\lambda_T$. 

We denote $\sum\limits_{i \in \bar{\mathcal{A}}}^{} \Vert \boldsymbol{\theta}^0_i - \tilde{\boldsymbol{\theta}}_i \Vert = \kappa_1$ and $\sum\limits_{i \in \bar{\mathcal{A}}^c}^{} \Vert \tilde{\boldsymbol{\theta}}_i \Vert = \kappa_2$. It holds that
\begin{eqnarray}
\notag &&\frac{1}{T} \Vert \boldsymbol{Z}_T (\boldsymbol{\theta}^0(T) - \tilde{\boldsymbol{\theta}}(T)) \Vert^2 \\
&\leq& \frac{2}{T} (\tilde{\boldsymbol{\theta}}(T) - \boldsymbol{\theta}^0(T))' \boldsymbol{Z}'_T u_T + \lambda_T \sum\limits_{i \in \bar{\mathcal{A}}}^{} \Vert \boldsymbol{\theta}^0_i - \tilde{\boldsymbol{\theta}}_i \Vert - \lambda_T \sum\limits_{i \in \bar{\mathcal{A}}^c}^{} \Vert \tilde{\boldsymbol{\theta}}_i \Vert \\
\notag &=& \frac{2}{T} (\tilde{\boldsymbol{\theta}}(T) - \boldsymbol{\theta}^0(T))' \boldsymbol{Z}'_T u_T + \lambda_T \kappa_1 - \lambda_T \kappa_2.
\label{eq:deltas}
\end{eqnarray}
Now, we consider two cases: $\kappa_2 > 2 \kappa_1$ and $\kappa_2 \leq 2 \kappa_1$. First, we show that $P(\kappa_2 > 2 \kappa_1) \to 0$ and then derive the upper bound of $\Vert \tilde{\boldsymbol{\theta}}_{\bar{\mathcal{A}}}(T) - \boldsymbol{\theta}^0_{\bar{\mathcal{A}}}(T) \Vert$.
We note that the triangle inequality yields
\begin{eqnarray}
\notag && \vert \frac{2}{T} (\tilde{\boldsymbol{\theta}}(T) - \boldsymbol{\theta}^0(T))' \boldsymbol{Z}'_T u_T \vert \\
&=& \frac{2}{T} \vert (\tilde{\boldsymbol{\theta}}_{\bar{\mathcal{A}}}(T) - \boldsymbol{\theta}^0_{\bar{\mathcal{A}}}(T))' \boldsymbol{Z}'_{T,\bar{\mathcal{A}}} u_T + (\tilde{\boldsymbol{\theta}}_{\bar{\mathcal{A}}^c}(T) - \boldsymbol{\theta}^0_{\bar{\mathcal{A}}^c}(T))' \boldsymbol{Z}'_{T,\bar{\mathcal{A}}^c} u_T \vert \\
\notag &\leq& \frac{2}{T} \vert (\tilde{\boldsymbol{\theta}}_{\bar{\mathcal{A}}}(T) - \boldsymbol{\theta}^0_{\bar{\mathcal{A}}}(T))' \boldsymbol{Z}'_{T,\bar{\mathcal{A}}} u_T \vert + \frac{2}{T} \vert (\tilde{\boldsymbol{\theta}}_{\bar{\mathcal{A}}^c}(T) - \boldsymbol{\theta}^0_{\bar{\mathcal{A}}^c}(T))' \boldsymbol{Z}'_{T,\bar{\mathcal{A}}^c} u_T \vert.
\label{eq:deltas_triangle}
\end{eqnarray}
We can show, using similar arguments as in Lemma 1, that $\underset{1 \leq i \leq T}{\max} \left( \sum\limits_{t=i}^{T} X_{jt}u_t \right)^2 = O_p(T^{5/2})$ and it follows that
\begin{eqnarray}
\Vert \boldsymbol{Z}'_{T,\bar{\mathcal{A}}} u_T \Vert^2 \leq N \underset{1 \leq j \leq N}{\max} \sum\limits_{i \in \bar{\mathcal{A}}}^{} \left( \sum\limits_{t=i}^{T} X_{jt} u_t \right)^2 \leq N \underset{1 \leq j \leq N}{\max} (m_0 + 1) \underset{1 \leq i \leq T}{\max} \left( \sum\limits_{t=i}^{T} X_{jt} u_t \right)^2.
\end{eqnarray}
Then, using the Cauchy-Schwarz inequality and considering that $|\bar{\mathcal{A}}| = m_0 + 1$ is finite, we have
\begin{eqnarray}
\frac{2}{T} \vert (\tilde{\boldsymbol{\theta}}_{\bar{\mathcal{A}}}(T) - \boldsymbol{\theta}^0_{\bar{\mathcal{A}}}(T))' \boldsymbol{Z}'_{T,\bar{\mathcal{A}}} u_T \vert \leq \frac{2}{T} \sum\limits_{i \in \bar{\mathcal{A}}}^{} \Vert \boldsymbol{\theta}^0_i - \tilde{\boldsymbol{\theta}}_i \Vert \Vert \boldsymbol{Z}'_{T,\bar{\mathcal{A}}} u_T \Vert = O_p(T^{1/4}) \kappa_1,
\end{eqnarray}
In contrast, $|\bar{\mathcal{A}}^c|$ is diverging with $T \to \infty$. We observe that 
\begin{eqnarray}
\notag \Vert \boldsymbol{Z}'_T u_T \Vert^2 &\leq& N \underset{1 \leq j \leq N}{\max} \sum\limits_{i=1}^{T-1} \left( \sum\limits_{t=i}^{T} X_{jt} u_t \right)^2 \leq N \underset{1 \leq j \leq N}{\max} T \underset{1 \leq i \leq T}{\max} \left( \sum\limits_{t=i}^{T} X_{jt} u_t \right)^2\\
&=& T O_p(T^{5/2}),
\end{eqnarray}
for all $j = 1, \dots, N$ which implies that $\Vert \boldsymbol{Z}'_T u_T \Vert = O_p(T^{7/4})$. Consequently, we have
\begin{equation}
\frac{2}{T} \vert (\tilde{\boldsymbol{\theta}}_{\bar{\mathcal{A}}^c}(T) - \boldsymbol{\theta}^0_{\bar{\mathcal{A}}^c}(T))' \boldsymbol{Z}'_{T,\bar{\mathcal{A}}^c} u_T \vert \leq \frac{2}{T} \sum\limits_{i \in \bar{\mathcal{A}}^c}^{} \Vert \tilde{\boldsymbol{\theta}}_i \Vert \Vert \boldsymbol{Z}'_{T,\bar{\mathcal{A}}^c} u_T \Vert = O_p(T^{3/4}) \kappa_2.
\end{equation}
Hence, using the assumption $\kappa_2 > 2 \kappa_1$, we obtain the contradiction
\begin{eqnarray}
\notag 0 &\leq&\frac{1}{T} \Vert \boldsymbol{Z}_T (\boldsymbol{\theta}^0(T) - \tilde{\boldsymbol{\theta}}(T)) \Vert^2 \\
&\leq& (O_p(T^{3/4}) - \lambda_T) \kappa_2 + (O_p(T^{1/4}) + \lambda_T) \kappa_1 \\
\notag &<& (O_p(T^{3/4}) + O_p(T^{1/4}) - \frac{1}{2} \lambda_T) \kappa_2 < 0,
\end{eqnarray}
for $T \to \infty$ and it follows that $P(\kappa_2 > 2 \kappa_1) \to 0$. Turning to the event $\kappa_2 \leq 2 \kappa_1$, Assumption \ref{as:2}(iii) and noting $\boldsymbol{X}' \boldsymbol{X}/T^2 = O_p(1)$ implies that
\begin{eqnarray}
\notag \Vert \boldsymbol{Z}_T (\boldsymbol{\theta}^0(T) - \tilde{\boldsymbol{\theta}}(T)) \Vert^2 &=& T^2 (\boldsymbol{\theta}^0(T) - \tilde{\boldsymbol{\theta}}(T))' \boldsymbol{\Sigma} (\boldsymbol{\theta}^0(T) - \tilde{\boldsymbol{\theta}}(T)) \\
&\geq& T^2 C \Vert \boldsymbol{\theta}^0_{\bar{\mathcal{A}}}(T) - \tilde{\boldsymbol{\theta}}_{\bar{\mathcal{A}}}(T) \Vert^2.
\end{eqnarray}
Thus, we have
\begin{eqnarray}
\Vert \boldsymbol{\theta}^0_{\bar{\mathcal{A}}}(T) - \tilde{\boldsymbol{\theta}}_{\bar{\mathcal{A}}}(T) \Vert &\leq& \sqrt{\frac{\Vert \boldsymbol{Z}_T (\boldsymbol{\theta}^0(T) - \tilde{\boldsymbol{\theta}}(T)) \Vert^2}{T^2 C}} \\
\notag &\leq& \sqrt{\frac{4 N c_0 T^{\delta+1} (m_0 + 1) M_{\beta}}{T^{2} C}} = \frac{2}{T^{(1-\delta)/2}} \sqrt{\frac{N c_0 (m_0 + 1) M_{\beta}}{C}} \to 0,
\end{eqnarray}
if $3/4 < \delta < 1$. Combining the results for both cases completes the proof.
\end{proofof}
$\hfill \Box$


\begin{proofof}
Define $A_{Ti} = \left\lbrace | \hat{t}_i - t^0_i | > T \epsilon \right\rbrace$, $i = 1,2, \dots, m_0$ such that
\begin{equation}
P \left( \underset{1 \leq i \leq m_0}{\max} | \hat{t}_i - t^0_i | > T \epsilon \right) \leq \sum\limits_{i=1}^{m_0} P \left( | \hat{t}_i - t^0_i | > T \epsilon \right) = \sum\limits_{i=1}^{m_0} P \left( A_{Ti} \right).
\end{equation}
Further define $C_T = \left\lbrace \underset{1 \leq i \leq m_0}{\max} | \hat{t}_i - t^0_i | \leq \underset{i}{\min} \, | t^0_i - t^0_{i-1} |/2 \right\rbrace$. It suffices to show that
\begin{equation}
\sum\limits_{i=1}^{m_0} P \left( A_{Ti} C_T \right) \to 0 \text{ and } \sum\limits_{i=1}^{m_0} P \left( A_{Ti} C^c_T \right) \to 0.
\end{equation}
The proof follows along the lines of the proof of Proposition 3 in \cite{Harchaoui2010} and Theorem 2.2 in \cite{ChanYauZhang2014}. In the following, we focus on $\sum\limits_{i=1}^{m_0} P \left( A_{Ti} C_T \right) \to 0$ because the complementary part can be shown using similar arguments.

In the set $C_T$, it holds that 
\begin{equation}
t^0_{i-1} < \hat{t}_i < t^0_{i+1}, \qquad \forall \, 1 \leq i \leq m_0. 
\end{equation}
Next, we split $A_{Ti}$ into two parts (i) $\hat{t}_i < t^0_i$ and (ii) $\hat{t}_i > t^0_i$ to show that $P \left( A_{Ti} C_T \right) \to 0$.

In case of (i), applying Lemma \ref{lemma:2} yields
\begin{equation}
\Vert \sum\limits_{s=\hat{t}_i}^{t^0_i-1} \boldsymbol{X}_s (y_s - \tilde{\boldsymbol{\beta}}_{i+1}' \boldsymbol{X}_s) \Vert \leq T \lambda_T.
\end{equation}
Note that because of $\hat{t}_i < t^0_i$, the true coefficient has not changed at $\hat{t}_i$. Hence, plugging in for $y_s = \boldsymbol{\beta}^{0'}_i \boldsymbol{X}_s + u_s$ yields
\begin{equation}
\Vert \sum\limits_{s=\hat{t}_i}^{t^0_i-1} \boldsymbol{X}_s u_s + \sum\limits_{s=\hat{t}_i}^{t^0_i-1} \boldsymbol{X}_s (\boldsymbol{\beta}^{0'}_i - \boldsymbol{\beta}^{0'}_{i+1}) \boldsymbol{X}_s + \sum\limits_{s=\hat{t}_i}^{t^0_i-1} \boldsymbol{X}_s (\boldsymbol{\beta}^{0'}_{i+1} - \tilde{\boldsymbol{\beta}}_{i+1}') \boldsymbol{X}_s \Vert \leq T \lambda_T.
\end{equation}
It follows for $\hat{t}_i < t^0_i$ that,
\begin{eqnarray}
P \left( A_{Ti} C_T \right) &\leq& P \left( \left\lbrace \frac{1}{3} \Vert \sum\limits_{s=\hat{t}_i}^{t^0_i-1} \boldsymbol{X}_s (\boldsymbol{\beta}^{0'}_i - \boldsymbol{\beta}^{0'}_{i+1}) \boldsymbol{X}_s \Vert \leq T \lambda_T \right\rbrace \cap \left\lbrace | \hat{t}_i - t^0_i | > T \epsilon \right\rbrace \right) \\
\notag &+& P \left( \left\lbrace \Vert \sum\limits_{s=\hat{t}_i}^{t^0_i-1} \boldsymbol{X}_s u_s \Vert > \frac{1}{3} \Vert \sum\limits_{s=\hat{t}_i}^{t^0_i-1} \boldsymbol{X}_s (\boldsymbol{\beta}^{0'}_i - \boldsymbol{\beta}^{0'}_{i+1}) \boldsymbol{X}_s \Vert \right\rbrace \cap \left\lbrace | \hat{t}_i - t^0_i | > T \epsilon \right\rbrace \right) \\
\notag &+& P \left( \left\lbrace \Vert \sum\limits_{s=\hat{t}_i}^{t^0_i-1} \boldsymbol{X}_s (\boldsymbol{\beta}^{0'}_{i+1} - \tilde{\boldsymbol{\beta}}_{i+1}') \boldsymbol{X}_s \Vert > \frac{1}{3} \Vert \sum\limits_{s=\hat{t}_i}^{t^0_i-1} \boldsymbol{X}_s (\boldsymbol{\beta}^{0'}_i - \boldsymbol{\beta}^{0'}_{i+1}) \boldsymbol{X}_s \Vert \right\rbrace \cap A_{Ti}C_T \right) \\
\notag &=& P \left( A_{Ti1} \right) + P \left( A_{Ti2} \right) + P \left( A_{Ti3} \right).
\end{eqnarray}
For the first term, we observe that under Assumption \ref{as:2} and on the set $\left\lbrace | \hat{t}_i - t^0_i | > T \epsilon \right\rbrace$ it holds that
\begin{equation}
\Vert \sum\limits_{s=\hat{t}_i}^{t^0_i-1} \boldsymbol{X}_s (\boldsymbol{\beta}^{0'}_i - \boldsymbol{\beta}^{0'}_{i+1}) \boldsymbol{X}_s \Vert \geq \Vert \sum\limits_{s=\hat{t}_i}^{t^0_i-1} \boldsymbol{X}_s \boldsymbol{X}_s' \Vert \underset{i}{\min} \left\lbrace \Vert \boldsymbol{\beta}^{0'}_i - \boldsymbol{\beta}^{0'}_{i+1} \Vert \right\rbrace  > \epsilon^2 \nu T^2,
\end{equation}
for sufficiently small $\epsilon$ with probability going to one. Taking into account that $T \lambda_T = O(T^{1+\delta})$ for $3/4 < \delta < 1$, we conclude that $P \left( A_{Ti1} \right) \to 0$ for $T \to \infty$.
For the second term, we have
\begin{equation}
\sum\limits_{s=\hat{t}_i}^{t^0_i-1} \boldsymbol{X}_s u_s = O_p (|t^0_i-1-\hat{t}_i|) = O_p(T),
\end{equation}
but since $\Vert \sum\limits_{s=\hat{t}_i}^{t^0_i-1} \boldsymbol{X}_s (\boldsymbol{\beta}^{0'}_i - \boldsymbol{\beta}^{0'}_{i+1}) \boldsymbol{X}_s \Vert > \epsilon^2 \nu T^2$ with probability going to one, we conclude that the right hand side of the inequality asymptotically dominates the left hand side and $P \left( A_{Ti2} \right) \to 0$ for $T \to \infty$. Turning to the third term, we note the definition of $\tilde{\boldsymbol{\beta}}_{i+1} = \sum\limits_{j=1}^{t} \tilde{\boldsymbol{\theta}}_j$ for $\hat{t}_{i} \leq t \leq \hat{t}_{i+1} - 1$, i.e., $\tilde{\boldsymbol{\beta}}_{i+1}$ is a linear function of the $\tilde{\boldsymbol{\theta}}_j$. Hence, according to Theorem \ref{th:1} and the Continuous Mapping Theorem (see \cite{Billingsley1999}, Theorem 2.7), $\tilde{\boldsymbol{\beta}}_{i+1}$ is a consistent estimator for $\boldsymbol{\beta}^0_{i+1}$ with convergence rate $T^{(1-\delta)/2}$, $3/4 < \delta < 1$. This means, we have $(\boldsymbol{\beta}^{0'}_{i+1} - \tilde{\boldsymbol{\beta}}_{i+1}') \to 0$ and $P \left( A_{Ti3} \right) \to 0$ for $T \to \infty$. It follows that $P \left( A_{Ti} C_T \cap \lbrace \hat{t}_i < t^0_i \rbrace \right) \to 0$.

In case of (ii), we have
\begin{equation}
\Vert \sum\limits_{s=t^0_i}^{\hat{t}_i-1} \boldsymbol{X}_s (y_s - \tilde{\boldsymbol{\beta}}_i' \boldsymbol{X}_s) \Vert \leq T \lambda_T,
\end{equation}

Since $\hat{t}_i > t^0_i$, the true coefficient has changed at $t^0_i$ and we plug in for $y_s = \boldsymbol{\beta}^{0'}_{i+1} \boldsymbol{X}_s + u_s$, which yields
\begin{equation}
\Vert \sum\limits_{s=t^0_i}^{\hat{t}_i-1} \boldsymbol{X}_s u_s + \sum\limits_{s=t^0_i}^{\hat{t}_i-1} \boldsymbol{X}_s (\boldsymbol{\beta}^{0'}_{i+1} - \boldsymbol{\beta}^{0'}_i) \boldsymbol{X}_s + \sum\limits_{s=t^0_i}^{\hat{t}_i-1} \boldsymbol{X}_s (\boldsymbol{\beta}^{0'}_i - \tilde{\boldsymbol{\beta}}_i') \boldsymbol{X}_s \Vert \leq T \lambda_T.
\end{equation}
It follows that,
\begin{eqnarray}
P \left( A_{Ti} C_T \right) &\leq& P \left( \left\lbrace \frac{1}{3} \Vert \sum\limits_{s=t^0_i}^{\hat{t}_i-1} \boldsymbol{X}_s (\boldsymbol{\beta}^{0'}_{i+1} - \boldsymbol{\beta}^{0'}_i) \boldsymbol{X}_s \Vert \leq T \lambda_T \right\rbrace \cap \left\lbrace | \hat{t}_i - t^0_i | > T \epsilon \right\rbrace \right) \\
\notag &+& P \left( \left\lbrace \Vert \sum\limits_{s=t^0_i}^{\hat{t}_i-1} \boldsymbol{X}_s u_s \Vert > \frac{1}{3} \Vert \sum\limits_{s=t^0_i}^{\hat{t}_i-1} \boldsymbol{X}_s (\boldsymbol{\beta}^{0'}_{i+1} - \boldsymbol{\beta}^{0'}_i) \boldsymbol{X}_s \Vert \right\rbrace \cap \left\lbrace | \hat{t}_i - t^0_i | > T \epsilon \right\rbrace \right) \\
\notag &+& P \left( \left\lbrace \Vert \sum\limits_{s=t^0_i}^{\hat{t}_i-1} \boldsymbol{X}_s (\boldsymbol{\beta}^{0'}_i - \tilde{\boldsymbol{\beta}}_i') \boldsymbol{X}_s \Vert > \frac{1}{3} \Vert \sum\limits_{s=t^0_i}^{\hat{t}_i-1} \boldsymbol{X}_s (\boldsymbol{\beta}^{0'}_{i+1} - \boldsymbol{\beta}^{0'}_i) \boldsymbol{X}_s \Vert \right\rbrace \cap A_{Ti}C_T \right) \\
\notag &=& P \left( A_{Ti1} \right) + P \left( A_{Ti2} \right) + P \left( A_{Ti3} \right).
\end{eqnarray}
The same arguments as for case (i) can be used to show that $P \left( A_{Ti} C_T \cap \lbrace \hat{t}_i > t^0_i \rbrace \right) \to 0$. Combining (i) and (ii) completes the proof of $P \left( A_{Ti} C_T \right) \to 0$.
\end{proofof}
$\hfill \Box$


\begin{proofof}
We begin to prove the first part. Suppose that $|\mathcal{A}_T| < m_0$, then there exists some $t^0_{i_0}$, $i_0 = 1,2, \dots$ and $\hat{t}_{s_0} \in \mathcal{A}_T \cup \lbrace 0, \infty \rbrace$, $s_0 = 0,1, \dots, |\mathcal{A}_T| + 1$ with $t^0_{i_0+1} - t^0_{i_0} \vee \hat{t}_{s_0} \geq T\epsilon/3$ and $t^0_{i_0+2} \wedge \hat{t}_{s_0+1} - t^0_{i_0+1} \geq T\epsilon/3$ where $\hat{t}_0 = 0$ and $\hat{t}_{|\mathcal{A}_T| + 1} = \infty$.


Applying Lemma \ref{lemma:2} to the intervals $[t^0_{i_0} \vee \hat{t}_{s_0}, t^0_{i_0+1}-1]$ and $[t^0_{i_0+1}, t^0_{i_0+2} \wedge \hat{t}_{s_0+1}-1]$ yields
\begin{equation}
\Vert \sum\limits_{s=t^0_{i_0} \vee \hat{t}_{s_0}}^{t^0_{i_0+1}-1} \boldsymbol{X}_s (y_s - \tilde{\boldsymbol{\beta}}_{s_0+1}' \boldsymbol{X}_s) \Vert \leq T \lambda_T,
\end{equation}
and
\begin{equation}
\Vert \sum\limits_{s=t^0_{i_0+1}}^{t^0_{i_0+2} \wedge \hat{t}_{s_0+1}-1} \boldsymbol{X}_s (y_s - \tilde{\boldsymbol{\beta}}_{s_0+1}' \boldsymbol{X}_s) \Vert \leq T \lambda_T.
\end{equation}
Similar arguments to those used in the proof of Theorem \ref{th:2} show that either 
\begin{eqnarray}
\notag && P \Bigg( \left\lbrace \Vert \sum\limits_{s=t^0_{i_0} \vee \hat{t}_{s_0}}^{t^0_{i_0+1}-1} \boldsymbol{X}_s (\boldsymbol{\beta}^{0'}_{i_0} - \tilde{\boldsymbol{\beta}}_{s_0+1}') \boldsymbol{X}_s \Vert > \frac{1}{3} \Vert \sum\limits_{s=t^0_{i_0} \vee \hat{t}_{s_0}}^{t^0_{i_0+1}-1} \boldsymbol{X}_s (\boldsymbol{\beta}^{0'}_{i_0} - \boldsymbol{\beta}^{0'}_{i_0+1}) \boldsymbol{X}_s \Vert \right\rbrace \\
&& \qquad \cap \lbrace |t^0_{i_0+1} - t^0_{i_0} \vee \hat{t}_{s_0}| \geq T\epsilon/3 \rbrace \Bigg) \to 0,
\label{eq:vee}
\end{eqnarray}
or
\begin{eqnarray}
\notag && P \Bigg( \left\lbrace \Vert \sum\limits_{s=t^0_{i_0+1}}^{t^0_{i_0+2} \wedge \hat{t}_{s_0+1}-1} \boldsymbol{X}_s (\boldsymbol{\beta}^{0'}_{i_0+1} - \tilde{\boldsymbol{\beta}}_{s_0+1}') \boldsymbol{X}_s \Vert > \frac{1}{3} \Vert \sum\limits_{s=t^0_{i_0+1}}^{t^0_{i_0+2} \wedge \hat{t}_{s_0+1}-1} \boldsymbol{X}_s (\boldsymbol{\beta}^{0'}_{i_0+1} - \boldsymbol{\beta}^{0'}_{i_0+2}) \boldsymbol{X}_s \Vert \right\rbrace \\
&& \qquad \cap \lbrace |t^0_{i_0+2} \wedge \hat{t}_{s_0+1} - t^0_{i_0+1}| \geq T\epsilon/3 \rbrace \Bigg) \to 0,
\label{eq:wedge}
\end{eqnarray}
has to hold to contradict $|\mathcal{A}_T| < m_0$. Since $\tilde{\boldsymbol{\beta}}_{s_0+1}$ is a consistent estimator according to Theorem \ref{th:1} and the Continuous Mapping Theorem, we either have $\tilde{\boldsymbol{\beta}}_{s_0+1} \overset{p}{\to} \boldsymbol{\beta}^0_{i_0}$ or $\tilde{\boldsymbol{\beta}}_{s_0+1} \overset{p}{\to} \boldsymbol{\beta}^0_{i_0+1}$. In the former case, the left hand side of Inequality~\eqref{eq:vee} converges to zero. In the latter case, the left hand side of Inequality~\eqref{eq:wedge} converges to zero. Hence, there is no situation in which not at least one probability converges to zero. Consequently, we have a contradiction to $|\mathcal{A}_T| < m_0$.

For the second part, we define $\hat{T}_k = \lbrace \hat{t}_1, \hat{t}_2, \dots, \hat{t}_{k} \rbrace$. Then, it is enough to show that
\begin{eqnarray}
P \left( \lbrace d_H(\mathcal{A}_T, \mathcal{A}) > T\epsilon, m_0 \leq |\mathcal{A}_T| \leq T \rbrace \right) \\
\notag = \sum\limits_{k=m_0}^{T} P \left( \lbrace d_H(\hat{T}_k, \mathcal{A}) > T\epsilon \rbrace \right) P \left( |\mathcal{A}_T| = k \right) \to 0,
\end{eqnarray}
as $T \to \infty$. By Theorem \ref{th:2}, we have already shown that $P \left( d_H(\hat{T}_{m_0}, \mathcal{A}) > T\epsilon \right) \to 0$ so that it suffices to show
\begin{equation}
\underset{k > m_0}{\max} \, P \left( d_H(\hat{T}_k, \mathcal{A}) > T\epsilon \right) \to 0.
\end{equation}
Given $t^0_i$, we define
\begin{eqnarray}
\notag B_{T,k,i,1} &=& \lbrace \forall 1 \leq s \leq k, |\hat{t}_s - t^0_i| \geq T\epsilon \text{ and } \hat{t}_s < t^0_i \rbrace \\
B_{T,k,i,2} &=& \lbrace \forall 1 \leq s \leq k, |\hat{t}_s - t^0_i| \geq T\epsilon \text{ and } \hat{t}_s > t^0_i \rbrace \\
\notag B_{T,k,i,3} &=& \lbrace \exists 1 \leq s \leq k - 1, \text{ such that } |\hat{t}_s - t^0_i| \geq T\epsilon, \\
\notag && \qquad |\hat{t}_{s+1} - t^0_i| \geq T\epsilon \text{ and } \hat{t}_s < t^0_i < \hat{t}_{s+1} \rbrace.
\end{eqnarray}
Then, 
\begin{equation}
\underset{k > m_0}{\max} \, P \left( d_H(\hat{T}_k, \mathcal{A}) > T\epsilon \right) = \underset{k > m_0}{\max} \, P \left( \bigcup\limits_{i=1}^{m_0} \bigcup\limits_{j=1}^{3} B_{T,k,i,j} \right).
\end{equation}
Using similar arguments as in the proof of Theorem \ref{th:2}, we can show that $\max_{k > m_0} \, P \left( \bigcup_{i=1}^{m_0} B_{T,k,i,j} \right) \to 0$ for $1 \leq j \leq 3$. This completes the proof of Theorem \ref{th:3}.
\end{proofof}
$\hfill \Box$


\begin{proofof}
To prove Theorem \ref{th:4}, we follow ideas similar to those put forth in \cite{WangLeng2008} and \cite{ZhangXiang2016}. As we will note at different points of the proof, the statistical properties of the adaptive group lasso estimator hinge crucially on our first step weights. It is particularly important that our second step design matrix $\boldsymbol{Z}_S$ fulfils the restricted eigenvalue condition which can be ensured by the first step group lasso algorithm.

We note that the adaptive group lasso objective function $Q(\boldsymbol{\theta}_S)$ is a strictly convex function and show that there is a local minimizer which is superconsistent. Then by global convexity of $Q(\boldsymbol{\theta}_S)$, it follows that such a local minimizer must be $\hat{\boldsymbol{\theta}}_S$. Similar as in \cite{FanLi2001}, the existence of an above-described local minimizer is implied by the fact that for any $\epsilon > 0$, there is a sufficiently large constant $C > 0$, such that
\begin{equation}
\underset{T}{\liminf} \, P \left( \underset{\boldsymbol{v} := (\boldsymbol{v}_1, \dots, \boldsymbol{v}_{M}) \in \mathbb{R}^{MN}: \Vert \boldsymbol{v} \Vert = C}{\inf} Q(\boldsymbol{\theta}^0_S + T^{-1} \boldsymbol{v}) > Q(\boldsymbol{\theta}^0_S) \right) > 1 - \epsilon.
\end{equation}
It holds that
\begin{eqnarray}
\notag && Q(\boldsymbol{\theta}^0_S + T^{-1} \boldsymbol{v}) - Q(\boldsymbol{\theta}^0_S) \\
\notag &=& \frac{1}{T} \Vert y_T - \boldsymbol{Z}_S (\boldsymbol{\theta}^0_S + T^{-1} \boldsymbol{v}) \Vert^2 + \lambda_S \sum\limits_{i=1}^{M} w_i \Vert (\boldsymbol{\theta}^0_{S,i} + T^{-1} \boldsymbol{v}_i) \Vert \\
\notag && - \frac{1}{T} \Vert y_T - \boldsymbol{Z}_S \boldsymbol{\theta}^0_S \Vert^2 - \lambda_S \sum\limits_{i=1}^{M} w_i \Vert \boldsymbol{\theta}^0_{S,i} \Vert \\
&=& \frac{1}{T} \boldsymbol{v}' \left( \frac{1}{T^2} \boldsymbol{Z}_S' \boldsymbol{Z}_S \right) \boldsymbol{v} - \frac{2}{T^{2}} \boldsymbol{v}' \boldsymbol{Z}_S' u_T \\
\notag && + \lambda_S \sum\limits_{i=1}^{M} w_i \Vert (\boldsymbol{\theta}^0_{S,i} + T^{-1} \boldsymbol{v}_i) \Vert - \lambda_S \sum\limits_{i=1}^{M} w_i \Vert \boldsymbol{\theta}^0_{S,i} \Vert \\
\notag &\geq& \frac{1}{T} \boldsymbol{v}' \left( \frac{1}{T^2} \boldsymbol{Z}_S' \boldsymbol{Z}_S \right) \boldsymbol{v} - \frac{2}{T^{2}} \boldsymbol{v}' \boldsymbol{Z}_S' u_T \\
\notag && + \lambda_S \sum\limits_{g(i) \in \mathcal{A}_T \cap \mathcal{A}}^{} \Vert \tilde{\boldsymbol{\theta}}_{S,i} \Vert^{-\gamma} \left( \Vert (\boldsymbol{\theta}^0_{S,i} + T^{-1} \boldsymbol{v}_i) \Vert - \Vert \boldsymbol{\theta}^0_{S,i} \Vert \right) \\
\notag &\geq& \frac{1}{T} \boldsymbol{v}' \left( \frac{1}{T^2} \boldsymbol{Z}_S' \boldsymbol{Z}_S \right) \boldsymbol{v} - \frac{2}{T^{2}} \boldsymbol{v}' \boldsymbol{Z}_S' u_T - \frac{1}{T} \lambda_S \sum\limits_{g(i) \in \mathcal{A}_T \cap \mathcal{A}}^{} \Vert \tilde{\boldsymbol{\theta}}_{S,i} \Vert^{-\gamma} \Vert \boldsymbol{v}_i \Vert \\
\notag &=& I_1 - I_2 - I_3.
\end{eqnarray}
Since the restricted eigenvalue condition holds for $\boldsymbol{\Sigma}_S = \boldsymbol{Z}_S' \boldsymbol{Z}_S / T^2$, i.e., its eigenvalues are positive for all $T$, and $\boldsymbol{\Sigma}_S$ thus converges to a positive definite random matrix, we have $I_1 = O_p(T^{-1}) \Vert \boldsymbol{v} \Vert^2$. Further, it follows from Cauchy-Schwarz inequality and Lemma \ref{lemma:1} that
\begin{eqnarray}
\notag E | I_2 |^2 &=& \frac{4}{T^{4}} E \left( \boldsymbol{v}' \boldsymbol{Z}_S' u_T \right)^2 \\
&\leq& \frac{4}{T^{4}} \Vert \boldsymbol{v} \Vert^2 E \Vert \boldsymbol{Z}_S' u_T \Vert^2 \\
\notag &=& \frac{1}{T^{2}} \Vert \boldsymbol{v} \Vert^2 O_p(1),
\end{eqnarray}
and consequently $I_2 = O_p(T^{-1}) \Vert \boldsymbol{v} \Vert$. Finally, using the Cauchy-Schwarz inequality, we have
\begin{eqnarray}
I_3 &\leq& \frac{1}{T} \lambda_S \left( \sum\limits_{g(i) \in \mathcal{A}_T \cap \mathcal{A}}^{} \Vert \tilde{\boldsymbol{\theta}}_{S,i} \Vert^{-2\gamma} \right)^{1/2} \Vert \boldsymbol{v} \Vert \\
\notag &\leq& \frac{1}{T} \lambda_S m_0^{1/2} \underset{g(i) \in \mathcal{A}_T \cap \mathcal{A}}{\min} \, \Vert \tilde{\boldsymbol{\theta}}_{S,i} \Vert^{-\gamma} \Vert \boldsymbol{v} \Vert.
\end{eqnarray}
We note that $\min_{g(i) \in \mathcal{A}_T \cap \mathcal{A}} \, \Vert \tilde{\boldsymbol{\theta}}_{S,i} \Vert^{-\gamma} = O_p(1)$ since $\tilde{\boldsymbol{\theta}}_{S,i}$ is a consistent estimator according to Theorem \ref{th:1} and our first step estimation does not ignore relevant breakpoints asymptotically according to Theorem \ref{th:3}. Using the condition $\lambda_S \to 0$, we know that $I_3$ is bounded by $O_p(T^{-1}) \Vert \boldsymbol{v} \Vert$. Hence, we can specify a large enough constant $C$ such that $I_1$ dominates $I_2$ and $I_3$. This completes the proof of part (a). 

Next, we turn to the proof of part (b). Lemma \ref{lemma:3} gives the necessary and sufficient condition for an estimator to be a solution to the adaptive group lasso objective function as defined by Equation~\eqref{eq:agfl}. Now, to prove that all truly zero parameters are set to zero almost surely, it suffices to show that
\begin{equation}
P \left( \forall g(i) \in \mathcal{A}_T \cap \mathcal{A}^c, \Vert \frac{1}{T} \boldsymbol{Z}_{g(i)}' (y_T - \boldsymbol{Z}_{S,\mathcal{A}^*} \hat{\boldsymbol{\theta}}_{S,\mathcal{A}^*}) \Vert \leq \frac{1}{2} \lambda_S \Vert \tilde{\boldsymbol{\theta}}_{S,i} \Vert^{-\gamma} \right) \to 1,
\end{equation}
or equivalently
\begin{equation}
P \left( \exists g(i) \in \mathcal{A}_T \cap \mathcal{A}^c, \Vert \frac{1}{T} \boldsymbol{Z}_{g(i)}' (y_T - \boldsymbol{Z}_{S,\mathcal{A}^*} \hat{\boldsymbol{\theta}}_{S,\mathcal{A}^*}) \Vert > \frac{1}{2} \lambda_S \Vert \tilde{\boldsymbol{\theta}}_{S,i} \Vert^{-\gamma} \right) \to 0,
\end{equation}
where $\boldsymbol{Z}_{g(i)}' = (0, \dots, 0, \boldsymbol{X}_{g(i)}', \boldsymbol{X}_{g(i)+1}',  \dots, \boldsymbol{X}_T')$.
It holds that
\begin{eqnarray}
&& P \left( \exists g(i) \in \mathcal{A}_T \cap \mathcal{A}^c, \Vert \frac{1}{T} \boldsymbol{Z}_{g(i)}' (y_T - \boldsymbol{Z}_{S,\mathcal{A}^*} \hat{\boldsymbol{\theta}}_{S,\mathcal{A}^*}) \Vert > \frac{1}{2} \lambda_S \Vert \tilde{\boldsymbol{\theta}}_{S,i} \Vert^{-\gamma} \right) \\
\notag &\leq& P \left( \exists g(i) \in \mathcal{A}_T \cap \mathcal{A}^c, \Vert \frac{1}{T} \boldsymbol{Z}_{g(i)}' u_T \Vert > \frac{1}{2} \lambda_S \Vert \tilde{\boldsymbol{\theta}}_{S,i} \Vert^{-\gamma} - \Vert \frac{1}{T} \boldsymbol{Z}_{g(i)}' \boldsymbol{Z}_{S,\mathcal{A}^*} \left( \hat{\boldsymbol{\theta}}_{S,\mathcal{A}^*} -  \boldsymbol{\theta}^0_{S,\mathcal{A}^*} \right) \Vert \right).
\end{eqnarray}
Further, we have
\begin{eqnarray}
\notag && \Vert \frac{1}{T} \boldsymbol{Z}_{g(i)}' \boldsymbol{Z}_{S,\mathcal{A}^*} \left( \hat{\boldsymbol{\theta}}_{S,\mathcal{A}^*} -  \boldsymbol{\theta}^0_{S,\mathcal{A}^*} \right) \Vert \\
&\leq& \left[ \left( \hat{\boldsymbol{\theta}}_{S,\mathcal{A}^*} -  \boldsymbol{\theta}^0_{S,\mathcal{A}^*} \right)' \boldsymbol{Z}_{S,\mathcal{A}^*}' \left( \frac{1}{T^2} \boldsymbol{Z}_{g(i)} \boldsymbol{Z}_{g(i)}' \right) \boldsymbol{Z}_{S,\mathcal{A}^*} \left( \hat{\boldsymbol{\theta}}_{S,\mathcal{A}^*} -  \boldsymbol{\theta}^0_{S,\mathcal{A}^*} \right) \right]^{1/2} \\
\notag &\leq& O_p(T) \Vert \hat{\boldsymbol{\theta}}_{S,\mathcal{A}^*} -  \boldsymbol{\theta}^0_{S,\mathcal{A}^*} \Vert,
\end{eqnarray}
and the first part of Theorem \ref{th:4} implies that $\Vert \hat{\boldsymbol{\theta}}_{S,\mathcal{A}^*} -  \boldsymbol{\theta}^0_{S,\mathcal{A}^*} \Vert = O_p(T^{-1})$ such that $\Vert \frac{1}{T} \boldsymbol{Z}_{g(i)}' \boldsymbol{Z}_{S,\mathcal{A}^*} \left( \hat{\boldsymbol{\theta}}_{S,\mathcal{A}^*} -  \boldsymbol{\theta}^0_{S,\mathcal{A}^*} \right) \Vert = O_p(1)$. Hence, we need to prove 
\begin{equation}
P \left( \exists g(i) \in \mathcal{A}_T \cap \mathcal{A}^c, \Vert \frac{1}{T} \boldsymbol{Z}_{g(i)}' u_T \Vert > \frac{1}{2} \lambda_S \Vert \tilde{\boldsymbol{\theta}}_{S,i} \Vert^{-\gamma} \right) \to 0.
\end{equation}
Considering that Theorem \ref{th:1} implies $\underset{g(i) \in \mathcal{A}^c}{\max} \, \Vert \tilde{\boldsymbol{\theta}}_{S,i} \Vert = O_p(T^{-(1-\delta)/2})$ for $3/4 < \delta < 1$, we have
\begin{eqnarray}
\notag && P \left( \exists g(i) \in \mathcal{A}_T \cap \mathcal{A}^c, \Vert \frac{1}{T} \boldsymbol{Z}_{g(i)}' u_T \Vert > \frac{1}{2} \lambda_S \Vert \tilde{\boldsymbol{\theta}}_{S,i} \Vert^{-\gamma} \right) \\
\notag &\leq& P \left( \exists g(i) \in \mathcal{A}_T \cap \mathcal{A}^c, \Vert \frac{1}{T} \boldsymbol{Z}_{g(i)}' u_T \Vert > \frac{\lambda_S}{ 2 \underset{i \in \mathcal{A}^c}{\max} \, \Vert \tilde{\boldsymbol{\theta}}_{S,i} \Vert^{\gamma}} \right) \\
&\leq& P \left( \exists g(i) \in \mathcal{A}_T \cap \mathcal{A}^c, \Vert \frac{1}{T} \boldsymbol{Z}_{g(i)}' u_T \Vert > \frac{1}{2} \lambda_S \left( C T^{-(1-\delta)/2} \right)^{-\gamma} \right) \\
\notag &\leq& \sum\limits_{g(i) \in \mathcal{A}_T \cap \mathcal{A}^c}^{} P \left( \Vert \frac{1}{T} \boldsymbol{Z}_{g(i)}' u_T \Vert > \frac{1}{2} \lambda_S \left( C T^{-(1-\delta)/2} \right)^{-\gamma} \right) \\
\notag &\leq& \sum\limits_{g(i) \in \mathcal{A}_T \cap \mathcal{A}^c}^{} \frac{4 E \Vert \frac{1}{T} \boldsymbol{Z}_{g(i)}' u_T \Vert^2}{\lambda^2_S \left( C T^{-(1-\delta)/2}, \right)^{-2\gamma}}
\end{eqnarray}
for some $C > 0$. Since Assumption \ref{as:1} implies $E \Vert \frac{1}{T} \boldsymbol{Z}_{g(i)}' u_T \Vert^2 = O_p(1)$ and $\lambda^2_S T^{(1-\delta)\gamma} \to \infty$, we have
\begin{equation}
\frac{4 E \Vert \frac{1}{T} \boldsymbol{Z}_{g(i)}' u_T \Vert^2}{\lambda^2_S C^{-2\gamma} T^{(1-\delta)\gamma}} \to 0,
\end{equation}
for all $g(i) \in \mathcal{A}_T \cap \mathcal{A}^c$. Note that $|\mathcal{A}_T| < M$ for all $T$ and that all remaining indices $i$ not included in $\mathcal{A}_T$ correspond to coefficients which have already been set to zero in the first step.

For the proof of model selection consistency, we still need to show that no truly non-zero parameter changes are set to zero. It holds that
\begin{equation}
\underset{g(i) \in \mathcal{A}}{\min} \, \Vert \hat{\boldsymbol{\theta}}_{S,i} \Vert \geq \underset{g(i) \in \mathcal{A}}{\min} \, \Vert \boldsymbol{\theta}^0_{S,i} \Vert - \underset{g(i) \in \mathcal{A}}{\max} \, \Vert \hat{\boldsymbol{\theta}}_{S,i} - \boldsymbol{\theta}^0_{S,i} \Vert.
\end{equation}
Since $\Vert \hat{\boldsymbol{\theta}}_{S,i} - \boldsymbol{\theta}^0_{S,i} \Vert \overset{p}{\to} 0$ by part (a) and by considering Assumption \ref{as:2}, we have
\begin{equation}
P \left( \underset{g(i) \in \mathcal{A}}{\min} \, \Vert \hat{\boldsymbol{\theta}}_{S,i} \Vert \geq \nu \right) \to 1. 
\end{equation}
This completes the proof of part (b).

Finally, we turn to the proof of part (c). It follows from Lemma \ref{lemma:3} that
\begin{equation}
P \left( - \frac{1}{T} \boldsymbol{Z}_{S,\bar{\mathcal{A}}^*}' \left( y_T - \boldsymbol{Z}_{S,\bar{\mathcal{A}}^*} \hat{\boldsymbol{\theta}}_{S,\bar{\mathcal{A}}^*} \right) + \frac{1}{2} \lambda_S \eta = 0 \right) \to 1,
\end{equation}
where
\begin{equation}
\eta = \left( \frac{\hat{\boldsymbol{\theta}}_{S,1}'}{\Vert \tilde{\boldsymbol{\theta}}_{S,1} \Vert^{\gamma} \Vert \hat{\boldsymbol{\theta}}_{S,1} \Vert}, \frac{\hat{\boldsymbol{\theta}}_{S,2}'}{\Vert \tilde{\boldsymbol{\theta}}_{S,2} \Vert^{\gamma} \Vert \hat{\boldsymbol{\theta}}_{S,2} \Vert}, \dots, \frac{\hat{\boldsymbol{\theta}}_{S,m_0+1}'}{\Vert \tilde{\boldsymbol{\theta}}_{S,m_0+1} \Vert^{\gamma} \Vert \hat{\boldsymbol{\theta}}_{S,m_0+1} \Vert} \right)'.
\end{equation}
Using $y_T = \boldsymbol{Z}_{S,\bar{\mathcal{A}}^*} \boldsymbol{\theta}^0_{S,\bar{\mathcal{A}}^*} + u_T$, we have
\begin{equation}
P \left( \frac{1}{T} \boldsymbol{Z}_{S,\bar{\mathcal{A}}^*}' \boldsymbol{Z}_{S,\bar{\mathcal{A}}^*} \left( \hat{\boldsymbol{\theta}}_{S,\bar{\mathcal{A}}^*} - \boldsymbol{\theta}^0_{S,\bar{\mathcal{A}}^*} \right) = \frac{1}{T} \boldsymbol{Z}_{S,\bar{\mathcal{A}}^*}' u_T - \frac{1}{2} \lambda_S \eta \right) \to 1.
\end{equation}
Then, it holds that
\begin{eqnarray}
\notag T \phi' (\hat{\boldsymbol{\theta}}_{S,\bar{\mathcal{A}}^*} - \boldsymbol{\theta}^0_{S,\bar{\mathcal{A}}^*}) &=& \frac{1}{T} \phi' \left( \frac{1}{T^2} \boldsymbol{Z}_{S,\bar{\mathcal{A}}^*}' \boldsymbol{Z}_{S,\bar{\mathcal{A}}^*} \right)^{-1}  \boldsymbol{Z}_{S,\bar{\mathcal{A}}^*}' u_T \\
	&-& \frac{1}{2} \lambda_S T \phi' \left( \frac{1}{T} \boldsymbol{Z}_{S,\bar{\mathcal{A}}^*}' \boldsymbol{Z}_{S,\bar{\mathcal{A}}^*} \right)^{-1} \eta + o_p(1),
\end{eqnarray}
where $\phi \in \mathbb{R}^{m_0 + 1}$ with $\Vert \phi \Vert = 1$. Since $\boldsymbol{Z}_{S,\bar{\mathcal{A}}^*}' \boldsymbol{Z}_{S,\bar{\mathcal{A}}^*}/ T^2$ is a positive definite random matrix for all $T$, we have 
\begin{equation}
\left\vert \frac{1}{2} \lambda_S \phi' \left( \frac{1}{T^2} \boldsymbol{Z}_{S,\bar{\mathcal{A}}^*}' \boldsymbol{Z}_{S,\bar{\mathcal{A}}^*} \right)^{-1} \eta \right\vert \leq \frac{1}{2} \lambda_S C \Vert \eta \Vert \Vert \phi \Vert \leq \frac{1}{2} \lambda_S C (m_0 + 1)^{1/2} \underset{i \in \bar{\mathcal{A}}}{\min} \, \Vert \tilde{\boldsymbol{\theta}}_i \Vert^{-\gamma}.
\end{equation}
As in part (a), it holds that $\min_{i \in \bar{\mathcal{A}}} \, \Vert \tilde{\boldsymbol{\theta}}_i \Vert^{-\gamma} = O_p(1)$ and by the conditions of Theorem 4, we have $\lambda_S \to 0$ as $T \to \infty$ such that
\begin{equation}
\left\vert \frac{1}{2} \lambda_S \phi' \left( \frac{1}{T^2} \boldsymbol{Z}_{S,\bar{\mathcal{A}}^*}' \boldsymbol{Z}_{S,\bar{\mathcal{A}}^*} \right)^{-1} \eta \right\vert = o_p(1).
\end{equation}
We use $\omega_{v_j}^2$ to denote the long-run variance of the stationary process $\lbrace v_{jt} \rbrace^{\infty}_{t=1}$, $j = 1, \dots, N$. Note that $\omega_{v_j}^2$ is the $j$-th diagonal element of   $\Omega_v$. Under the conditions of Assumption \ref{as:1}, it holds that
\begin{equation}
T^{-1/2} X_{j[Ts]} = T^{-1/2} \sum\limits_{t=1}^{\lbrack Ts \rbrack} v_{jt} \Rightarrow B(s),
\label{eq:fclt}
\end{equation}
for $s \in \lbrack 0, 1 \rbrack$, $j \in \lbrace 1, \dots, N \rbrace$ and $T \to \infty$, where $B(s)$ is a scalar Brownian motion with variance $\omega_{v_j}^2$. Further, it holds that
\begin{equation}
T^{-1/2} \boldsymbol{Z}_{S,\mathcal{A}^*,[Ts]} \Rightarrow (\boldsymbol{B}(s), \boldsymbol{B}(s) \varphi_{\tau_1}(s), \dots, \boldsymbol{B}(s) \varphi_{\tau_{m_0}}(s)) \equiv \boldsymbol{B}_{\tau, \bar{\mathcal{A}}^*}(s),
\end{equation}
where
\begin{equation}
\varphi_{\tau_k}(s) \begin{cases} \hphantom{-} 0 & \text{ if } s < \tau_k \\ \hphantom{-} 1 & \text{ if } s \geq \tau_k \end{cases}, \qquad k \in \lbrace 1, \dots, m_0 \rbrace, \quad s \in [0,1],
\end{equation}
and $\boldsymbol{B}(s)$ is $N$-vector Brownian motion process with covariance matrix $\Omega_v$. Using (A.4) in \cite{GregoryHansen1996} and the Continuous Mapping Theorem, we observe that
\begin{equation}
\frac{1}{T^2} \boldsymbol{Z}_{S,\bar{\mathcal{A}}^*}' \boldsymbol{Z}_{S,\bar{\mathcal{A}}^*} \Rightarrow \int\limits_{0}^{1} \boldsymbol{B}_{\tau, \bar{\mathcal{A}}^*} (s)'  \boldsymbol{B}_{\tau, \bar{\mathcal{A}}^*} (s) ds,
\end{equation}
where the weak convergence is uniform over the vector $(\tau_1, \dots, \tau_{m_0}) \in \mathcal{T}$. Further, using (A.3) in \cite{GregoryHansen1996} and Theorem 3.1 in \cite{Hansen1992b}, we have the weak convergence to a stochastic integral
\begin{equation}
\frac{1}{T} \boldsymbol{Z}_{S,\bar{\mathcal{A}}^*}' u_T \Rightarrow \int\limits_{0}^{1} \boldsymbol{B}_{\tau, \bar{\mathcal{A}}^*} (s) d U (s) + \begin{bmatrix}
    \Lambda \\
    (1 - \tau_1) \Lambda \\
    \vdots \\
    (1 - \tau_{m_0}) \Lambda
        \end{bmatrix},
\end{equation}
where $\Lambda = \sum\limits_{t=0}^{\infty} E (v_0 u_t)$. Finally, the Cram\'{e}r-Wold device implies the weak convergence result in (c) which completes the proof of Theorem 4.
\end{proofof}
$\hfill \Box$

\clearpage

\begin{landscape}

\section{Tables}\label{sec:tables}

\begin{table}[htbp]
\caption{Estimation of (multiple) structural breaks in the slope coefficients (Adaptive Group Lasso)}
\begin{center}
\scalebox{0.55}{
\begin{tabular}{c c c c c c c c c c c c c c c c c c c c c c c c c c c c c c}
\toprule[1pt]
 &  \multicolumn{16}{l}{SB1: $\mu = 2$, $\boldsymbol{\theta}_k = 2$, $k = \lbrace 1, 2 \rbrace$, ($\tau = 0.5$)} \\ 
$T$ & & $pce$ &  & $hd/T$ &  & $\tau$ &  & $\theta_{1,1}$ &  & $\theta_{1,2}$ &  & $\theta_{2,1}$ &  & $\theta_{2,2}$\\ 
\midrule[0.5pt]
100 & & 100 & & 0.48 & & 0.501 (0.022) & & 2.01 (0.144) & & 1.99 (0.168) & & 2.01 (0.155) & & 1.98 (0.168)\\
200 & & 100 & & 0.17 & & 0.500 (0.009) & & 2.01 (0.073) & & 2.00 (0.087) & & 2.00 (0.073) & & 1.99 (0.086)\\
400 & & 100 & & 0.08 & & 0.500 (0.004) & & 2.00 (0.042) & & 2.00 (0.057) & & 2.00 (0.040) & & 2.00 (0.053)\\
 & & & & & & & & & & & & & & \\ 
 &  \multicolumn{16}{l}{SB2: $\mu = 2$, $\boldsymbol{\theta}_k = 2$, $k = \lbrace 1, 2, 3 \rbrace$, ($\tau_1 = 0.33$, $\tau_2 = 0.67$)} \\ 
$T$ & & $pce$ &  & $hd/T$ &  & $\tau_1$ &  & $\tau_2$ &  & $\theta_{1,1}$ &  & $\theta_{1,2}$ &  & $\theta_{1,3}$ &  & $\theta_{2,1}$ &  & $\theta_{2,2}$ &  & $\theta_{2,3}$ \\
\midrule[0.5pt]
100 & & 99.8 &  & 0.69 &  & 0.327 (0.024) &  & 0.672 (0.024) &  & 2.01 (0.221) &  & 2.00 (0.254) &  & 1.98 (0.306) &  & 2.02 (0.217) &  & 2.01 (0.269) &  & 1.97 (0.321) \\
200 & & 100 &  & 0.25 &  & 0.329 (0.010) &  & 0.670 (0.006) &  & 2.00 (0.092) &  & 2.00 (0.117) &  & 2.00 (0.114) &  & 2.01 (0.093) &  & 2.00 (0.116) &  & 2.00 (0.105) \\
400 & & 100 &  & 0.06 &  & 0.330 (0.004) &  & 0.670 (0.002) &  & 2.00 (0.047) &  & 2.00 (0.059) &  & 2.00 (0.053) &  & 2.00 (0.046) &  & 2.00 (0.058) &  & 2.00 (0.055) \\
 & & & & & & & & & & & & & & \\ 
 &  \multicolumn{16}{l}{SB4: $\mu = 2$, $\boldsymbol{\theta}_k = 2$, $k = \lbrace 1, \dots, 5 \rbrace$, ($\tau_1 = 0.2$, $\tau_2 = 0.4$, $\tau_3 = 0.6$, $\tau_4 = 0.8$)} \\ 
$T$ & & $pce$ &  & $hd/T$ &  & $\tau_1$ &  & $\tau_2$ &  & $\tau_3$ &  & $\tau_4$\\
\midrule[0.5pt]
100 & & 96.5 & & 1.66 & & 0.201 (0.024) & & 0.402 (0.032) & & 0.601 (0.026) & & 0.801 (0.016)\\
200 & & 100 & & 0.57 & & 0.199 (0.008) & & 0.400 (0.006) & & 0.600 (0.006) & & 0.800 (0.009)\\
400 & & 100 & & 0.16 & & 0.200 (0.006) & & 0.400 (0.003) & & 0.600 (0.002) & & 0.800 (0.003)\\
 & & & & & & & & & & & & & & \\
$T$ & & $\theta_{1,1}$ &  & $\theta_{1,2}$ &  & $\theta_{1,3}$ &  & $\theta_{1,4}$ &  & $\theta_{1,5}$\\ 
\midrule[0.5pt]
100 & & 2.04 (0.380) & & 2.01 (0.497) & & 1.98 (0.544) & & 1.96 (0.556) & & 1.99 (0.504) \\
200 & & 2.02 (0.191) & & 1.99 (0.204) & & 2.00 (0.213) & & 2.00 (0.203) & & 2.00 (0.203) \\
400 & & 2.00 (0.074) & & 2.00 (0.096) & & 2.00 (0.092) & & 2.00 (0.087) & & 2.00 (0.091) \\
 & & & & & & & & & & & & & & \\
$T$ & & $\theta_{2,1}$ &  & $\theta_{2,2}$ &  & $\theta_{2,3}$ &  & $\theta_{2,4}$ &  & $\theta_{2,5}$\\
\midrule[0.5pt]
100 & & 2.04 (0.358) & & 2.01 (0.426) & & 1.96 (0.495) & & 1.97 (0.532) & & 1.99 (0.447) \\
200 & & 2.02 (0.211) & & 1.99 (0.222) & & 2.00 (0.208) & & 1.99 (0.203) & & 2.00 (0.203) \\
400 & & 2.00 (0.074) & & 2.00 (0.097) & & 2.00 (0.090) & & 2.00 (0.087) & & 2.00 (0.087) \\
\bottomrule[1pt]
\end{tabular}
}
\end{center}
\label{tab:sb_diverg_sig2}
\begin{tablenotes}
\scriptsize
\item Note: We use 1,000 replications of the data-generating process given in Equation~\eqref{eq:mc.dgp}. The variance of the error terms is $\sigma^2_{\omega} = 1$ and $\sigma^2_{\vartheta} = 4$, respectively. The first panel reports the results for one active breakpoint at $\tau = 0.5$, the second panel considers two active breakpoints at $\tau_1 = 0.33$ and $\tau_2 = 0.67$ and the third panel has four active breakpoints at $\tau_1 = 0.2$, $\tau_2 = 0.4$, $\tau_3 = 0.6$, and $\tau_4 = 0.8$. The baseline coefficients and parameter changes at all breakpoints take the value 2. Standard deviations are given in parentheses.
\end{tablenotes}
\end{table}
\end{landscape}

\begin{landscape}
\begin{table}[htbp]
\caption{Estimation of (multiple) structural breaks in the slope coefficients (Bai-Perron)}
\begin{center}
\scalebox{0.55}{
\begin{tabular}{c c c c c c c c c c c c c c c c c c c c c c c c c c c c c c}
\toprule[1pt]
 &  \multicolumn{16}{l}{SB1: $\mu = 2$, $\boldsymbol{\theta}_k = 2$, $k = \lbrace 1, 2 \rbrace$, ($\tau = 0.5$)} \\ 
$T$ & & $pce$ &  & $hd/T$ &  & $\tau$ &  & $\theta_{1,1}$ &  & $\theta_{1,2}$ &  & $\theta_{2,1}$ &  & $\theta_{2,2}$\\ 
\midrule[0.5pt]
100 & & 100 & & 0.23 & & 0.500 (0.009) & & 2.00 (0.155) & & 2.00 (0.214) & & 2.00 (0.156) & & 1.99 (0.207)\\
200 & & 100 & & 0.07 & & 0.500 (0.004) & & 2.00 (0.077) & & 2.00 (0.109) & & 2.00 (0.076) & & 2.00 (0.106)\\
400 & & 100 & & 0.03 & & 0.500 (0.002) & & 2.00 (0.039) & & 2.00 (0.054) & & 2.00 (0.040) & & 2.00 (0.054)\\
 & & & & & & & & & & & & & & \\ 
 &  \multicolumn{16}{l}{SB2: $\mu = 2$, $\boldsymbol{\theta}_k = 2$, $k = \lbrace 1, 2, 3 \rbrace$, ($\tau_1 = 0.33$, $\tau_2 = 0.67$)} \\ 
$T$ & & $pce$ &  & $hd/T$ &  & $\tau_1$ &  & $\tau_2$ &  & $\theta_{1,1}$ &  & $\theta_{1,2}$ &  & $\theta_{1,3}$ &  & $\theta_{2,1}$ &  & $\theta_{2,2}$ &  & $\theta_{2,3}$ \\
\midrule[0.5pt]
100 & & 100 &  & 0.40 &  & 0.330 (0.008) &  & 0.670 (0.008) &  & 2.00 (0.238) &  & 2.01 (0.322) &  & 1.99 (0.338) &  & 2.01 (0.239) &  & 1.99 (0.338) &  & 1.99 (0.327) \\
200 & & 100 &  & 0.15 &  & 0.330 (0.004) &  & 0.670 (0.003) &  & 2.00 (0.120) &  & 1.99 (0.165) &  & 2.00 (0.164) &  & 2.00 (0.119) &  & 2.01 (0.165) &  & 2.00 (0.159) \\
400 & & 100 &  & 0.06 &  & 0.330 (0.002) &  & 0.670 (0.002) &  & 2.00 (0.059) &  & 2.00 (0.080) &  & 2.00 (0.082) &  & 2.00 (0.058) &  & 2.00 (0.078) &  & 2.00 (0.082) \\
 & & & & & & & & & & & & & & \\ 
 &  \multicolumn{16}{l}{SB4: $\mu = 2$, $\boldsymbol{\theta}_k = 2$, $k = \lbrace 1, \dots, 5 \rbrace$, ($\tau_1 = 0.2$, $\tau_2 = 0.4$, $\tau_3 = 0.6$, $\tau_4 = 0.8$)} \\ 
$T$ & & $pce$ &  & $hd/T$ &  & $\tau_1$ &  & $\tau_2$ &  & $\tau_3$ &  & $\tau_4$\\
\midrule[0.5pt]
100 & & 96.7 & & 0.86 & & 0.200 (0.011) & & 0.401 (0.009) & & 0.600 (0.011) & & 0.799 (0.011)\\
200 & & 100 & & 0.32 & & 0.200 (0.005) & & 0.400 (0.004) & & 0.600 (0.004) & & 0.800 (0.004)\\
400 & & 100 & & 0.13 & & 0.200 (0.002) & & 0.400 (0.002) & & 0.600 (0.001) & & 0.800 (0.002)\\
 & & & & & & & & & & & & & & \\
$T$ & & $\theta_{1,1}$ &  & $\theta_{1,2}$ &  & $\theta_{1,3}$ &  & $\theta_{1,4}$ &  & $\theta_{1,5}$\\ 
\midrule[0.5pt]
100 & & 1.97 (0.400) & & 2.05 (0.558) & & 2.00 (0.550) & & 2.00 (0.587) & & 1.98 (0.575) \\
200 & & 2.00 (0.194) & & 1.99 (0.266) & & 2.00 (0.266) & & 2.00 (0.279) & & 1.98 (0.271) \\
400 & & 2.00 (0.096) & & 2.00 (0.134) & & 2.00 (0.133) & & 2.00 (0.135) & & 2.00 (0.132) \\
 & & & & & & & & & & & & & & \\
$T$ & & $\theta_{2,1}$ &  & $\theta_{2,2}$ &  & $\theta_{2,3}$ &  & $\theta_{2,4}$ &  & $\theta_{2,5}$\\
\midrule[0.5pt]
100 & & 2.02 (0.434) & & 2.00 (0.539) & & 1.98 (0.559) & & 1.99 (0.556) & & 2.01 (0.559) \\
200 & & 2.00 (0.202) & & 2.00 (0.274) & & 2.01 (0.251) & & 2.00 (0.272) & & 2.02 (0.270) \\
400 & & 2.00 (0.095) & & 2.00 (0.136) & & 2.00 (0.135) & & 2.00 (0.133) & & 2.01 (0.136) \\
\bottomrule[1pt]
\end{tabular}
}
\end{center}
\label{tab:sb_diverg_sig2_baiperron}
\begin{tablenotes}
\scriptsize
\item Note: We use 1,000 replications of the data-generating process given in Equation~\eqref{eq:mc.dgp}. The variance of the error terms is $\sigma^2_{\omega} = 1$ and $\sigma^2_{\vartheta} = 4$, respectively. The first panel reports the results for one active breakpoint at $\tau = 0.5$, the second panel considers two active breakpoints at $\tau_1 = 0.33$ and $\tau_2 = 0.67$ and the third panel has four active breakpoints at $\tau_1 = 0.2$, $\tau_2 = 0.4$, $\tau_3 = 0.6$, and $\tau_4 = 0.8$. The baseline coefficients and parameter changes at all breakpoints take the value 2. Standard deviations are given in parentheses.
\end{tablenotes}
\end{table}
\end{landscape}

\begin{landscape}
\begin{table}[htbp]
\caption{Estimation of partial structural breaks}
\begin{center}
\scalebox{0.55}{
\begin{tabular}{c c c c c c c c c c c c c c c c c c c c c c c c c c c c c c}
\toprule[1pt]
 &  \multicolumn{16}{l}{SB1: $\mu = 2$, $\boldsymbol{\theta}_1 = 2$, $\theta_{2,1} = 2$, $\theta_{2,2} = 0$ ($\tau = 0.5$)} \\ 
$T$ & & $pce$ &  & $hd/T$ &  & $\tau$ &  & $\theta_{1,1}$ &  & $\theta_{1,2}$ &  & $\theta_{2,1}$ &  & $\theta_{2,2}$\\ 
\midrule[0.5pt]
100 & & 99.6 & & 0.80 & & 0.499 (0.033) & & 2.01 (0.145) & & 1.98 (0.186) & & 2.00 (0.137) & & $-0.01$ (0.158)\\
200 & & 100 & & 0.26 & & 0.501 (0.013) & & 2.00 (0.078) & & 1.99 (0.104) & & 2.00 (0.074) & & 0.00 (0.082)\\
400 & & 100 & & 0.09 & & 0.500 (0.005) & & 2.00 (0.040) & & 2.00 (0.053) & & 2.00 (0.034) & & 0.00 (0.042)\\
 & & & & & & & & & & & & & & \\ 
 &  \multicolumn{16}{l}{SB2: $\mu = 2$, $\boldsymbol{\theta}_1 = 2$, $\theta_{2,1} = 2$, $\theta_{2,j} = 0$, $j = \lbrace 2, 3 \rbrace$, ($\tau_1 = 0.33$, $\tau_2 = 0.67$)} \\ 
$T$ & & $pce$ &  & $hd/T$ &  & $\tau_1$ &  & $\tau_2$ &  & $\theta_{1,1}$ &  & $\theta_{1,2}$ &  & $\theta_{1,3}$ &  & $\theta_{2,1}$ &  & $\theta_{2,2}$ &  & $\theta_{2,3}$ \\
\midrule[0.5pt]
100 & & 98.2 &  & 1.58 &  & 0.328 (0.034) &  & 0.672 (0.032) &  & 2.03 (0.221) &  & 1.99 (0.270) &  & 1.97 (0.293) &  & 2.00 (0.190) &  & $-0.01$ (0.230) &  & 0.00 (0.226) \\
200 & & 100 &  & 0.57 &  & 0.328 (0.012) &  & 0.671 (0.015) &  & 2.01 (0.105) &  & 1.99 (0.134) &  & 2.00 (0.130) &  & 2.00 (0.092) &  & 0.00 (0.117) &  & 0.00 (0.105) \\
400 & & 100 &  & 0.22 &  & 0.330 (0.005) &  & 0.670 (0.005) &  & 2.00 (0.049) &  & 2.00 (0.091) &  & 2.00 (0.078) &  & 2.00 (0.047) &  & 0.00 (0.058) &  & 0.00 (0.055) \\
 & & & & & & & & & & & & & & \\ 
 &  \multicolumn{16}{l}{SB4: $\mu = 2$, $\boldsymbol{\theta}_1 = 2$, $\theta_{2,1} = 2$, $\theta_{2,j} = 0$, $j = \lbrace 2, \dots, 5 \rbrace$, ($\tau_1 = 0.2$, $\tau_2 = 0.4$, $\tau_3 = 0.6$, $\tau_4 = 0.8$)} \\ 
$T$ & & $pce$ &  & $hd/T$ &  & $\tau_1$ &  & $\tau_2$ &  & $\tau_3$ &  & $\tau_4$\\
\midrule[0.5pt]
100 & & 94.9 & & 2.86 & & 0.205 (0.035) & & 0.401 (0.037) & & 0.600 (0.040) & & 0.797 (0.033)\\
200 & & 99.4 & & 1.10 & & 0.200 (0.015) & & 0.400 (0.013) & & 0.600 (0.015) & & 0.800 (0.015)\\
400 & & 100 & & 0.37 & & 0.200 (0.005) & & 0.400 (0.004) & & 0.600 (0.004) & & 0.800 (0.005)\\
 & & & & & & & & & & & & & & \\
$T$ & & $\theta_{1,1}$ &  & $\theta_{1,2}$ &  & $\theta_{1,3}$ &  & $\theta_{1,4}$ &  & $\theta_{1,5}$\\ 
\midrule[0.5pt]
100 & & 2.07 (0.408) & & 2.00 (1.436) & & 1.93 (1.527) & & 1.96 (0.691) & & 1.98 (0.539) \\
200 & & 2.03 (0.217) & & 1.99 (0.245) & & 1.98 (0.276) & & 2.00 (0.236) & & 2.00 (0.211) \\
400 & & 2.01 (0.085) & & 1.99 (0.104) & & 2.00 (0.101) & & 2.00 (0.097) & & 2.00 (0.096) \\
 & & & & & & & & & & & & & & \\
$T$ & & $\theta_{2,1}$ &  & $\theta_{2,2}$ &  & $\theta_{2,3}$ &  & $\theta_{2,4}$ &  & $\theta_{2,5}$\\
\midrule[0.5pt]
100 & & 2.01 (0.313) & & $-0.02$ (0.533) & & $-0.01$ (0.527) & & 0.00 (0.502) & & 0.01 (0.448) \\
200 & & 2.00 (0.154) & & 0.00 (0.188) & & 0.00 (0.183) & & 0.00 (0.187) & & 0.01 (0.177) \\
400 & & 2.00 (0.074) & & 0.00 (0.094) & & 0.00 (0.089) & & 0.00 (0.087) & & 0.01 (0.087) \\
\bottomrule[1pt]
\end{tabular}
}
\end{center}
\label{tab:sb_partial_sig2}
\begin{tablenotes}
\scriptsize
\item Note: We use 1,000 replications of the data-generating process given in Equation~\eqref{eq:mc.dgp}. The variance of the error terms is $\sigma^2_{\omega} = 1$ and $\sigma^2_{\vartheta} = 4$, respectively. The first panel reports the results for one active breakpoint at $\tau = 0.5$, the second panel considers two active breakpoints at $\tau_1 = 0.33$ and $\tau_2 = 0.67$ and the third panel has four active breakpoints at $\tau_1 = 0.2$, $\tau_2 = 0.4$, $\tau_3 = 0.6$, and $\tau_4 = 0.8$. The baseline coefficients and take the value 2. We induce partial structural breaks through a change in $\theta_1$ only. Standard deviations are given in parentheses.
\end{tablenotes}
\end{table}
\end{landscape}

\begin{landscape}
\begin{table}[htbp]
\caption{Break fractions located near the boundaries of the unit interval}
\begin{center}
\scalebox{0.55}{
\begin{tabular}{c c c c c c c c c c c c c c c c c c c c c c c c c c c c c c}
\toprule[1pt]
 &  \multicolumn{16}{l}{SB1: $\mu = 2$, $\boldsymbol{\theta}_k = 2$, $k = \lbrace 1, 2 \rbrace$, ($\tau = 0.1$)} \\ 
$T$ & & $pce$ &  & $hd/T$ &  & $\tau$ &  & $\theta_{1,1}$ &  & $\theta_{1,2}$ &  & $\theta_{2,1}$ &  & $\theta_{2,2}$\\ 
\midrule[0.5pt]
100 & & 89.0 & & 0.19 & & 0.100 (0.013) & & 2.01 (0.622) & & 2.00 (0.621) & & 2.04 (0.581) & & 1.96 (0.584)\\
200 & & 96.6 & & 0.18 & & 0.098 (0.007) & & 2.00 (0.300) & & 2.00 (0.302) & & 2.01 (0.276) & & 1.99 (0.276)\\
400 & & 99.7 & & 0.15 & & 0.099 (0.007) & & 2.01 (0.150) & & 2.00 (0.151) & & 2.00 (0.142) & & 2.00 (0.142)\\
 & & & & & & & & & & & & & & \\
 &  \multicolumn{16}{l}{SB1: $\mu = 2$, $\boldsymbol{\theta}_k = 2$, $k = \lbrace 1, 2 \rbrace$, ($\tau = 0.9$)} \\ 
$T$ & & $pce$ &  & $hd/T$ &  & $\tau$ &  & $\theta_{1,1}$ &  & $\theta_{1,2}$ &  & $\theta_{2,1}$ &  & $\theta_{2,2}$\\ 
\midrule[0.5pt]
100 & & 95.2 & & 0.21 & & 0.901 (0.008) & & 2.01 (0.090) & & 1.98 (0.477) & & 2.00 (0.081) & & 1.94 (0.493)\\
200 & & 98.0 & & 0.18 & & 0.901 (0.008) & & 2.00 (0.042) & & 1.99 (0.239) & & 2.00 (0.044) & & 2.00 (0.242)\\
400 & & 99.5 & & 0.08 & & 0.901 (0.005) & & 2.00 (0.022) & & 1.99 (0.132) & & 2.00 (0.021) & & 2.00 (0.124)\\
 & & & & & & & & & & & & & & \\ 
 &  \multicolumn{16}{l}{SB2: $\mu = 2$, $\boldsymbol{\theta}_k = 2$, $k = \lbrace 1, 2, 3 \rbrace$, ($\tau_1 = 0.1$, $\tau_2 = 0.9$)} \\ 
$T$ & & $pce$ &  & $hd/T$ &  & $\tau_1$ &  & $\tau_2$ &  & $\theta_{1,1}$ &  & $\theta_{1,2}$ &  & $\theta_{1,3}$ &  & $\theta_{2,1}$ &  & $\theta_{2,2}$ &  & $\theta_{2,3}$ \\
\midrule[0.5pt]
100 & & 86.4 &  & 0.50 &  & 0.100 (0.016) &  & 0.901 (0.013) &  & 2.01 (0.623) &  & 2.00 (0.624) &  & 1.98 (0.490) &  & 2.04 (0.587) &  & 1.95 (0.589) &  & 1.95 (0.480) \\
200 & & 94.8 &  & 0.47 &  & 0.099 (0.017) &  & 0.901 (0.010) &  & 2.00 (0.278) &  & 2.00 (0.280) &  & 1.98 (0.248) &  & 2.02 (0.287) &  & 1.98 (0.288) &  & 1.99 (0.244) \\
400 & & 99.3 &  & 0.28 &  & 0.099 (0.008) &  & 0.901 (0.005) &  & 2.01 (0.158) &  & 1.99 (0.159) &  & 2.00 (0.132) &  & 2.01 (0.147) &  & 1.99 (0.146) &  & 2.00 (0.123) \\
 & & & & & & & & & & & & & & \\ 
 &  \multicolumn{16}{l}{SB2: $\mu = 2$, $\boldsymbol{\theta}_k = 2$, $k = \lbrace 1, 2, 3 \rbrace$, ($\tau_1 = 0.1$, $\tau_2 = 0.95$)} \\ 
$T$ & & $pce$ &  & $hd/T$ &  & $\tau_1$ &  & $\tau_2$ &  & $\theta_{1,1}$ &  & $\theta_{1,2}$ &  & $\theta_{1,3}$ &  & $\theta_{2,1}$ &  & $\theta_{2,2}$ &  & $\theta_{2,3}$ \\
\midrule[0.5pt]
100 & & 83.9 &  & 0.49 &  & 0.102 (0.023) &  & 0.950 (0.012) &  & 2.02 (0.640) &  & 1.99 (0.638) &  & 1.95 (0.876) &  & 2.05 (0.597) &  & 1.95 (0.599) &  & 1.98 (0.855) \\
200 & & 92.8 &  & 0.40 &  & 0.098 (0.008) &  & 0.950 (0.018) &  & 2.00 (0.270) &  & 2.00 (0.270) &  & 1.97 (0.513) &  & 2.02 (0.275) &  & 1.98 (0.277) &  & 1.97 (0.516) \\
400 & & 97.7 &  & 0.26 &  & 0.099 (0.007) &  & 0.951 (0.005) &  & 2.01 (0.154) &  & 1.99 (0.154) &  & 1.99 (0.251) &  & 2.01 (0.143) &  & 2.00 (0.143) &  & 2.00 (0.244) \\
\bottomrule[1pt]
\end{tabular}
}
\end{center}
\label{tab:sb_bounds_sig2}
\begin{tablenotes}
\scriptsize
\item Note: We use 1,000 replications of the data-generating process given in Equation~\eqref{eq:mc.dgp}. The variance of the error terms is $\sigma^2_{\omega} = 1$ and $\sigma^2_{\vartheta} = 4$, respectively. The first panel reports the results for one active breakpoint near the left boundary at $\tau = 0.1$, the second panel considers one active breakpoint near the right boundary at $\tau = 0.9$. The third panel has two active breakpoints at $\tau_1 = 0.1$ and $\tau_2 = 0.9$ and the fourth panel features breakpoints at $\tau_1 = 0.1$ and $\tau_2 = 0.95$. The baseline coefficients and parameter changes at all breakpoints take the value 2. Standard deviations are given in parentheses.
\end{tablenotes}
\end{table}
\end{landscape}

\begin{table}[htbp]
\caption{Endogeneity correction via dynamic augmentation}
\begin{center}
\scalebox{0.5}{
\begin{tabular}{c c c c c c c c c c c c c c c c c c c c c c c c c c c c c c}
\toprule[1pt]
 &  \multicolumn{16}{l}{SB1: $\mu = 2$, $\boldsymbol{\theta}_k = 2$, $k = \lbrace 1, 2 \rbrace$, ($\tau = 0.1$)} \\ 
$T$ & & $pce$ &  & $hd/T$ &  & $\tau$ &  & $\theta_{1,1}$ &  & $\theta_{1,2}$ &  & $\theta_{2,1}$ &  & $\theta_{2,2}$\\ 
\midrule[0.5pt]
 & & & & & & $l=1$\\
\cmidrule[0.5pt]{7-7}
100 & & 99.9 & & 0.54 & & 0.500 (0.025) & & 2.00 (0.153) & & 1.99 (0.170) & & 2.01 (0.147) & & 1.99 (0.170)\\
200 & & 100 & & 0.14 & & 0.500 (0.008) & & 2.00 (0.075) & & 2.00 (0.088) & & 2.00 (0.069) & & 2.00 (0.084)\\
 & & & & & & $l=2$\\
\cmidrule[0.5pt]{7-7}
100 & & 99.9 & & 0.48 & & 0.500 (0.022) & & 2.00 (0.169) & & 1.99 (0.201) & & 2.01 (0.170) & & 1.99 (0.212)\\
200 & & 100 & & 0.16 & & 0.500 (0.009) & & 2.00 (0.075) & & 2.00 (0.087) & & 2.00 (0.071) & & 2.00 (0.085)\\
 & & & & & & & & & & & & & & \\
 &  \multicolumn{16}{l}{SB2: $\mu = 2$, $\boldsymbol{\theta}_k = 2$, $k = \lbrace 1, 2, 3 \rbrace$, ($\tau_1 = 0.33$, $\tau_2 = 0.67$)} \\ 
$T$ & & $pce$ &  & $hd/T$ &  & $\tau_1$ &  & $\tau_2$ &  & $\theta_{1,1}$ &  & $\theta_{1,2}$ &  & $\theta_{1,3}$ &  & $\theta_{2,1}$ &  & $\theta_{2,2}$ &  & $\theta_{2,3}$ \\
\midrule[0.5pt]
 & & & & & & $l=1$\\
\cmidrule[0.5pt]{7-7}
100 & & 99.2 & & 0.93 &  & 0.329 (0.025) &  & 0.672 (0.019) &  & 2.03 (0.292) &  & 1.97 (0.329) &  & 1.99 (0.274) &  & 2.03 (0.285) &  & 1.98 (0.328) &  & 1.98 (0.284) \\
200 & & 100 & & 0.35 &  & 0.330 (0.008) &  & 0.670 (0.010) &  & 2.00 (0.099) &  & 2.00 (0.128) &  & 2.00 (0.141) &  & 2.01 (0.134) &  & 2.00 (0.128) &  & 2.00 (0.158) \\
 & & & & & & $l=2$\\
\cmidrule[0.5pt]{7-7}
100 & & 99.4 & & 1.06 &  & 0.329 (0.029) &  & 0.670 (0.026) &  & 2.03 (0.290) &  & 1.97 (0.348) &  & 1.99 (0.283) &  & 2.03 (0.299) &  & 1.97 (0.364) &  & 1.99 (0.310) \\
200 & & 100 & & 0.37 &  & 0.329 (0.008) &  & 0.671 (0.011) &  & 2.00 (0.101) &  & 2.00 (0.135) &  & 2.00 (0.123) &  & 2.00 (0.109) &  & 2.00 (0.138) &  & 2.00 (0.133) \\
 & & & & & & & & & & & & & & \\ 
 &  \multicolumn{16}{l}{SB4: $\mu = 2$, $\boldsymbol{\theta}_1 = 2$, $\theta_{2,1} = 2$, $\theta_{2,j} = 0$, $j = \lbrace 2, \dots, 5 \rbrace$, ($\tau_1 = 0.2$, $\tau_2 = 0.4$, $\tau_3 = 0.6$, $\tau_4 = 0.8$)} \\ 
$T$ & & $pce$ &  & $hd/T$ &  & $\tau_1$ &  & $\tau_2$ &  & $\tau_3$ &  & $\tau_4$\\
\midrule[0.5pt]
 & & & & & & $l=1$\\
\cmidrule[0.5pt]{7-7}
100 & & 98.3 & & 1.42 & & 0.203 (0.024) & & 0.402 (0.023) & & 0.600 (0.022) & & 0.800 (0.017)\\
200 & & 99.4 & & 0.59 & & 0.199 (0.010) & & 0.400 (0.006) & & 0.600 (0.009) & & 0.800 (0.009)\\
 & & & & & & $l=2$\\
\cmidrule[0.5pt]{7-7}
100 & & 97.9 & & 1.52 & & 0.202 (0.022) & & 0.401 (0.023) & & 0.600 (0.023) & & 0.799 (0.019)\\
200 & & 99.7 & & 0.64 & & 0.199 (0.011) & & 0.400 (0.008) & & 0.600 (0.011) & & 0.800 (0.011)\\
 & & & & & & & & & & & & & & \\
$T$ & & $\theta_{1,1}$ &  & $\theta_{1,2}$ &  & $\theta_{1,3}$ &  & $\theta_{1,4}$ &  & $\theta_{1,5}$ & & $\theta_{2,1}$ &  & $\theta_{2,2}$ &  & $\theta_{2,3}$ &  & $\theta_{2,4}$ &  & $\theta_{2,5}$\\ 
\midrule[0.5pt]
 & & & & & & $l=1$\\
\cmidrule[0.5pt]{7-7}
100 & & 2.04 (0.386) & & 2.00 (0.422) & & 1.95 (0.487) & & 1.98 (0.457) & & 2.01 (0.417) & & 2.05 (0.403) & & 1.99 (0.476) & & 1.99 (0.503) & & 1.95 (0.474) & & 1.99 (0.417) \\
200 & & 2.01 (0.160) & & 2.00 (0.190) & & 1.99 (0.185) & & 2.01 (0.258) & & 1.99 (0.258) & & 2.01 (0.157) & & 2.01 (0.193) & & 1.99 (0.197) & & 2.01 (0.261) & & 1.99 (0.257) \\
 & & & & & & $l=2$\\
 \cmidrule[0.5pt]{7-7}
100 & & 2.04 (0.397) & & 2.00 (0.415) & & 2.01 (0.460) & & 1.97 (0.452) & & 2.02 (0.373) & & 2.05 (0.409) & & 1.98 (0.437) & & 2.01 (0.480) & & 1.94 (0.485) & & 2.01 (0.380) \\
200 & & 2.01 (0.191) & & 2.00 (0.195) & & 1.98 (0.242) & & 2.01 (0.203) & & 1.99 (0.181) & & 2.01 (0.174) & & 2.01 (0.199) & & 1.98 (0.261) & & 2.00 (0.215) & & 1.99 (0.181) \\
\bottomrule[1pt]
\end{tabular}
}
\end{center}
\label{tab:sb_endogenous_sig2}
\begin{tablenotes}
\scriptsize
\item Note: We use 1,000 replications of the data-generating process given in Equation~\eqref{eq:mc.dgp} with an endogenous error term specification. The covariance matrix of the error terms is specified according to Equation~\eqref{eq:V} with $\sigma^2_{\omega} = 1$ and $\sigma^2_{\vartheta} = 4$, respectively. We denote the number of leads and lags with $l$. The first panel reports the results for one active breakpoint at $\tau = 0.5$, the second panel considers two active breakpoints at $\tau_1 = 0.33$ and $\tau_2 = 0.67$ and the third panel has four active breakpoints at $\tau_1 = 0.2$, $\tau_2 = 0.4$, $\tau_3 = 0.6$, and $\tau_4 = 0.8$. The baseline coefficients and parameter changes at all breakpoints take the value 2. Standard deviations are given in parentheses.
\end{tablenotes}
\end{table}

\clearpage

\bibliographystyle{elsarticle-harv}
\bibliography{ci_agl_bibfile}

\begin{thebibliography}{73}
\expandafter\ifx\csname natexlab\endcsname\relax\def\natexlab#1{#1}\fi
\expandafter\ifx\csname url\endcsname\relax
  \def\url#1{\texttt{#1}}\fi
\expandafter\ifx\csname urlprefix\endcsname\relax\def\urlprefix{URL }\fi

\bibitem[{Andrews(1993)}]{Andrews1993}
Andrews, D. W.~K., 1993. {Tests for Parameter Instability and Structural Change
  With Unknown Change Point}. Econometrica 61~(4), 821--856.

\bibitem[{Arai and Kurozumi(2007)}]{AraiKurozumi2007}
Arai, Y., Kurozumi, E., 2007. {Testing for the Null Hypothesis of Cointegration
  with a Structural Break}. Econometric Reviews 26~(6), 705--739.

\bibitem[{Aue and Horv{\'{a}}th(2013)}]{AueHorvath2013}
Aue, A., Horv{\'{a}}th, L., 2013. {Structural breaks in time series}. Journal
  of Time Series Analysis 34~(1), 1--16.

\bibitem[{Bae and Jong(2007)}]{Bae2007}
Bae, Y., Jong, R. M.~D., 2007. {Money Demand Function Estimation by Nonlinear
  Cointegration}. Journal of Applied Econometrics 22, 767--793.

\bibitem[{Bai et~al.(1998)Bai, Lumsdaine, and Stock}]{BaiLumsdaineStock1998}
Bai, J., Lumsdaine, R.~L., Stock, J.~H., 1998. {Testing for and Dating Common
  Breaks in Multivariate Time Series}. The Review of Economic Studies 65~(3),
  395--432.

\bibitem[{Bai and Perron(1998)}]{BaiPerron1998}
Bai, J., Perron, P., 1998. {Estimating and Testing Linear Models with Multiple
  Structural Changes}. Econometrica 66~(1), 47--78.

\bibitem[{Bai and Perron(2003)}]{BaiPerron2003}
Bai, J., Perron, P., 2003. {Computation and analysis of multiple structural
  change models}. Journal of Applied Econometrics 18~(1), 1--22.

\bibitem[{Bai and Perron(2006)}]{BaiPerron2006}
Bai, J., Perron, P., 2006. {Multiple structural change models: A simulation
  analysis}. In: Corbae, D., Durlauf, S., Hansen, B. (Eds.), Econometric Theory
  and Practice. Cambridge University Press, pp. 212--337.

\bibitem[{Ball(2001)}]{Ball2001}
Ball, L., 2001. {Another look at long-run money demand}. Journal of Monetary
  Economics 47~(1), 31--44.

\bibitem[{Behrendt and Schweikert(2020)}]{Behrendt2020}
Behrendt, S., Schweikert, K., 2020. {A Note on Adaptive Group Lasso for
  Structural Break Time Series}. Econometrics and Statistics, forthcoming.

\bibitem[{Bickel et~al.(2009)Bickel, Ritov, and
  Tsybakov}]{BickelRitovTsybakov2009}
Bickel, P.~J., Ritov, Y., Tsybakov, A.~B., 2009. {Simultaneous analysis of
  lasso and dantzig selector}. Annals of Statistics 37~(4), 1705--1732.

\bibitem[{Billingsley(1999)}]{Billingsley1999}
Billingsley, P., 1999. Convergence of Probability Measures, 2nd Edition. Wiley,
  New York.

\bibitem[{Boot and Pick(2019)}]{BootPick2019}
Boot, T., Pick, A., 2019. {Does modeling a structural break improve forecast
  accuracy?} Journal of Econometrics, 1--25.

\bibitem[{Boysen et~al.(2009)Boysen, Kempe, Liebscher, Munk, and
  Wittich}]{Boysen2009}
Boysen, L., Kempe, A., Liebscher, V., Munk, A., Wittich, O., 2009.
  {Consistencies and rates of convergence of jump-penalized least squares
  estimators}. Annals of Statistics 37~(1), 157--183.

\bibitem[{Breheny and Huang(2009)}]{BrehenyHuang2009}
Breheny, P., Huang, J., 2009. {Penalized methods for bi-level variable
  selection.} Statistics and its interface 2~(3), 369--380.

\bibitem[{Campos et~al.(1996)Campos, Ericsson, and Hendry}]{Campos1996}
Campos, J., Ericsson, N.~R., Hendry, D.~F., 1996. {Cointegration tests in the
  presence of structural breaks}. Journal of Econometrics 70~(1), 187--220.

\bibitem[{Carrion-i Silvestre and Sanso(2006)}]{Carrion-i-Silvestre2006}
Carrion-i Silvestre, J.~L., Sanso, A., 2006. {Testing for the Null Hypothesis
  of Cointegration with Structural Breaks}. Oxford Bulletin of Economics and
  Statistics 68~(5), 623--646.

\bibitem[{Chan et~al.(2014)Chan, Yau, and Zhang}]{ChanYauZhang2014}
Chan, N.~H., Yau, C.~Y., Zhang, R.-M., 2014. {Group LASSO for Structural Break
  Time Series}. Journal of the American Statistical Association 109~(506),
  590--599.

\bibitem[{Chen and Wu(2005)}]{ChenWu2005}
Chen, S.~L., Wu, J.~L., 2005. {Long-run money demand revisited: Evidence from a
  non-linear approach}. Journal of International Money and Finance 24~(1),
  19--37.

\bibitem[{Ciuperca(2014)}]{Ciuperca2014}
Ciuperca, G., 2014. {Model selection by LASSO methods in a change-point model}.
  Statistical Papers 55~(2), 349--374.

\bibitem[{Davidson and Monticini(2010)}]{DavidsonMonticini2010}
Davidson, J., Monticini, A., 2010. {Tests for cointegration with structural
  breaks based on subsamples}. Computational Statistics and Data Analysis
  54~(11), 2498--2511.

\bibitem[{Davis et~al.(2006)Davis, Lee, and Rodriguez-Yam}]{Davis2006}
Davis, R.~A., Lee, T. C.~M., Rodriguez-Yam, G.~A., 2006. {Structural Break
  Estimation for Nonstationary Time Series Models}. Journal of the American
  Statistical Association 101~(473), 223--239.

\bibitem[{Fan and Li(2001)}]{FanLi2001}
Fan, J., Li, R., 2001. {Variable selection via nonconcave penalized likelihood
  and its oracle properties}. Journal of the American Statistical Association
  96~(456), 1348--1360.

\bibitem[{Gregory and Hansen(1996{\natexlab{a}})}]{GregoryHansen1996}
Gregory, A.~W., Hansen, B.~E., 1996{\natexlab{a}}. {Residual-based tests for
  cointegration in models with regime shifts}. Journal of Econometrics 70~(2),
  99--126.

\bibitem[{Gregory and Hansen(1996{\natexlab{b}})}]{GregoryHansen1996b}
Gregory, A.~W., Hansen, B.~E., 1996{\natexlab{b}}. {Tests for Cointegration in
  Models with Regime and Trend Shifts}. Oxford Bulletin of Economics and
  Statistics 58~(3), 555--560.

\bibitem[{Gregory et~al.(1996)Gregory, Nason, and Watt}]{GregoryNasonWatt1996}
Gregory, A.~W., Nason, J.~M., Watt, D.~G., 1996. {Testing for structural breaks
  in cointegrated relationships}. Journal of Econometrics 71~(1-2), 321--341.

\bibitem[{Hansen(1992{\natexlab{a}})}]{Hansen1992a}
Hansen, B.~E., 1992{\natexlab{a}}. {Convergence to Stochastic Integrals for
  Dependent Heterogeneous Processes}. Econometric Theory 8~(4), 489--500.

\bibitem[{Hansen(1992{\natexlab{b}})}]{Hansen1992b}
Hansen, B.~E., 1992{\natexlab{b}}. {Efficient estimation and testing of
  cointegrating vectors in the presence of deterministic trends}. Journal of
  Econometrics 53, 87--121.

\bibitem[{Harchaoui and L{\'{e}}vy-Leduc(2010)}]{Harchaoui2010}
Harchaoui, Z., L{\'{e}}vy-Leduc, C., 2010. {Multiple Change-Point Estimation
  With a Total Variation Penalty}. Journal of the American Statistical
  Association 105~(April), 1480--1493.

\bibitem[{Hatemi-J(2008)}]{Hatemi-J2008}
Hatemi-J, A., 2008. {Tests for cointegration with two unknown regime shifts
  with an application to financial market integration}. Empirical Economics
  35~(3), 497--505.

\bibitem[{He and Huang(2016)}]{HeHuang2016}
He, K., Huang, J.~Z., 2016. {Asymptotic properties of adaptive group Lasso for
  sparse reduced rank regression}. Stat 5~(1), 251--261.

\bibitem[{Hoffman and Rasche(1991)}]{HoffmanRasche1991}
Hoffman, D.~L., Rasche, R.~H., 1991. {Long-Run Income and Interest Elasticities
  of Money Demand in the United States}. The Review of Economics and Statistics
  73~(4), 665--674.

\bibitem[{Horowitz and Huang(2013)}]{HorowitzHuang2013}
Horowitz, J., Huang, J., 2013. {Penalized estimation of high-dimensional models
  under a generalized sparsity condition}. Statistica Sinica 23~(2), 725--748.

\bibitem[{Huang et~al.(2009)Huang, Ma, Xie, and Zhang}]{HuangMaXieZhang2009}
Huang, J., Ma, S., Xie, H., Zhang, C.~H., 2009. {A group bridge approach for
  variable selection}. Biometrika 96~(2), 339--355.

\bibitem[{Huang et~al.(2008)Huang, Ma, and Zhang}]{HuangMaZhang2008}
Huang, J., Ma, S., Zhang, C.-H., 2008. {Adaptive Lasso for Sparse
  High-Dimensional Regression Models}. Statistica Sinica 18, 1603--1618.

\bibitem[{Ireland(2009)}]{Ireland2009}
Ireland, P.~N., 2009. {On the Welfare Cost of Inflation and the Recent Behavior
  of Money Demand}. The American Economic Review 99~(3), 1040--1052.

\bibitem[{Jawadi and Sousa(2013)}]{JawadiSousa2013}
Jawadi, F., Sousa, R.~M., 2013. {Money demand in the euro area, the US and the
  UK: Assessing the role of nonlinearity}. Economic Modelling 32~(1), 507--515.

\bibitem[{Jin et~al.(2013)Jin, Shi, and Wu}]{JinShiWu2013}
Jin, B., Shi, X., Wu, Y., 2013. {A novel and fast methodology for simultaneous
  multiple structural break estimation and variable selection for nonstationary
  time series models}. Statistics and Computing 23~(2), 221--231.

\bibitem[{Jin et~al.(2016)Jin, Wu, and Shi}]{JinWuShi2016}
Jin, B., Wu, Y., Shi, X., 2016. {Consistent two-stage multiple change-point
  detection in linear models}. Canadian Journal of Statistics 44~(2), 161--179.

\bibitem[{Johansen(2005)}]{Johansen2005}
Johansen, S., 2005. {Interpretation of cointegrating coefficients in the
  cointegrated vector autoregressive model}. Oxford Bulletin of Economics and
  Statistics 67~(1), 93--104.

\bibitem[{Juselius(2006)}]{Juselius2006}
Juselius, K., 2006. The Cointegrated VAR Model: Methodology and Applications.
  Oxford University Press, United Kingdom.

\bibitem[{Kejriwal and Perron(2008)}]{KejriwalPerron2008}
Kejriwal, M., Perron, P., 2008. {The limit distribution of the estimates in
  cointegrated regression models with multiple structural changes}. Journal of
  Econometrics 146~(1), 59--73.

\bibitem[{Kejriwal and Perron(2010)}]{KejriwalPerron2010}
Kejriwal, M., Perron, P., 2010. {Testing for Multiple Structural Changes in
  Cointegrated Regression Models}. Journal of Business {\&} Economic Statistics
  28~(4), 503--522.

\bibitem[{Kock(2016)}]{Kock2016}
Kock, A.~B., 2016. {Consistent and Conservative Model Selection With the
  Adaptive Lasso in Stationary and Nonstationary Autoregressions}. Econometric
  Theory 32~(1), 243--259.

\bibitem[{Li and Perron(2017)}]{LiPerron2017}
Li, Y., Perron, P., 2017. {Inference on locally ordered breaks in multiple
  regressions}. Econometric Reviews 36~(1-3), 289--353.

\bibitem[{Lucas(1988)}]{Lucas1988}
Lucas, R.~E., 1988. {Money Demand in the United States: A Quantitive Review}.
  Carnegie-Rochester Conference on Public Policy 29, 137--168.

\bibitem[{Lucas(2000)}]{Lucas2000}
Lucas, R.~E., 2000. {Inflation and Welfare}. Econometrica 68~(2), 247--274.

\bibitem[{Lucas and Nicolini(2015)}]{LucasNicolini2015}
Lucas, R.~E., Nicolini, J.~P., 2015. {On the stability of money demand}.
  Journal of Monetary Economics 73, 48--65.

\bibitem[{Maki(2012)}]{Maki2012}
Maki, D., 2012. {Tests for cointegration allowing for an unknown number of
  breaks}. Economic Modelling 29~(5), 2011--2015.

\bibitem[{Meinshausen and B{\"{u}}hlmann(2006)}]{Meinshausen2006}
Meinshausen, N., B{\"{u}}hlmann, P., 2006. {High-dimensional graphs and
  variable selection with the Lasso}. Annals of Statistics 34~(3), 1436--1462.

\bibitem[{Meltzer(1963)}]{Meltzer1963}
Meltzer, A.~H., 1963. {The Demand for Money: The Evidence from the Time
  Series}. The Journal of Political Economy 71~(3), 219--246.

\bibitem[{Mogliani and Urga(2018)}]{MoglianiUrga2018}
Mogliani, M., Urga, G., 2018. {On the Instability of Long-Run Money Demand and
  the Welfare Cost of Inflation in the United States}. Journal of Money, Credit
  and Banking 50~(7), 1645--1660.

\bibitem[{Niu et~al.(2015)Niu, Hao, and Zhang}]{NiuHaoZhang2015}
Niu, Y.~S., Hao, N., Zhang, H., 2015. {Multiple Change-point Detection: A
  Selective Overview}. Statistical Science 31~(4), 611--623.

\bibitem[{Oka and Perron(2018)}]{OkaPerron2018}
Oka, T., Perron, P., 2018. {Testing for common breaks in a multiple equations
  system}. Journal of Econometrics 204~(1), 66--85.

\bibitem[{Perron(2006)}]{Perron2006}
Perron, P., 2006. {Dealing with structural breaks}. In: Hassani, H., Mills, T.,
  Patterson, K. (Eds.), Palgrave Handbook of Econometrics - Volume 1:
  Econometric Theory. Palgrave Macmillan UK, pp. 278--352.

\bibitem[{Qian and Jia(2016)}]{QianJia2016}
Qian, J., Jia, J., 2016. {On stepwise pattern recovery of the fused Lasso}.
  Computational Statistics and Data Analysis 94, 221--237.

\bibitem[{Qian and Su(2016)}]{QianSu2016}
Qian, J., Su, L., 2016. {Shrinkage Estimation of Regression Models With
  Multiple Structural Changes}. Econometric Theory 32~(6), 1376--1433.

\bibitem[{Qu(2007)}]{Qu2007}
Qu, Z., 2007. {Searching for cointegration in a dynamic system}. Econometrics
  Journal 10~(3), 580--604.

\bibitem[{Qu and Perron(2007)}]{QuPerron2007}
Qu, Z., Perron, P., 2007. {Estimating and testing structural change in
  multivariate regressions}. Econometrica 75~(2), 459--502.

\bibitem[{Saikkonen(1991)}]{Saikkonen1991}
Saikkonen, P., 1991. Asymptotically efficient estimation of cointegration
  regressions. Econometric Theory 7~(1), 1--21.

\bibitem[{Schmidt and Schweikert(2019)}]{SchmidtSchweikert2019}
Schmidt, A., Schweikert, K., 2019. {Multiple Structural Breaks in Cointegrating
  Regressions: A Model Selection Approach}. SSRN Workingpaper, 1--49.
\newline\urlprefix\url{https://ssrn.com/abstract=3489870}

\bibitem[{Schweikert(2020)}]{Schweikert2019}
Schweikert, K., 2020. {Testing for cointegration with threshold adjustment in
  the presence of structural breaks}. Studies in Nonlinear Dynamics \&
  Econometrics 24~(1), 1--26.

\bibitem[{Stock and Watson(1993)}]{StockWatson1993}
Stock, J.~H., Watson, M.~W., 1993. {A simple estimator of cointegrating vectors
  in higher order integrated systems}. Econometrica 61~(4), 783--820.

\bibitem[{Teles and Zhou(2005)}]{TelesZhou2005}
Teles, P., Zhou, R., 2005. {A stable money demand: Looking for the right
  monetary aggregate}. Economic Perspectives~(Q I), 50--63.

\bibitem[{Wang and Leng(2008)}]{WangLeng2008}
Wang, H., Leng, C., 2008. {A note on adaptive group lasso}. Computational
  Statistics and Data Analysis 52~(12), 5277--5286.

\bibitem[{Wang et~al.(2009)Wang, Li, and Leng}]{WangLiLeng2009}
Wang, H., Li, B., Leng, C., 2009. {Shrinkage tuning parameter selection with a
  diverging number of parameters}. Journal of the Royal Statistical Society.
  Series B: Statistical Methodology 71~(3), 671--683.

\bibitem[{Wang(2011)}]{Wang2011}
Wang, Y., 2011. {The stability of long-run money demand in the United States: A
  new approach}. Economics Letters 111~(1), 60--63.

\bibitem[{Wei and Huang(2010)}]{WeiHuang2010}
Wei, F., Huang, J., 2010. {Consistent group selection in high-dimensional
  linear regression}. Bernoulli 16~(4), 1369--1384.

\bibitem[{Westerlund and Edgerton(2006)}]{WesterlundEdgerton2006}
Westerlund, J., Edgerton, D.~L., 2006. {New Improved Tests for Cointegration
  with Structural Breaks}. Journal of Time Series Analysis 28~(2), 188--224.

\bibitem[{Yuan and Lin(2006)}]{YuanLin2006}
Yuan, M., Lin, Y., 2006. {Model selection and estimation in regression with
  grouped variables}. Journal of the Royal Statistical Society. Series B:
  Statistical Methodology 68~(1), 49--67.

\bibitem[{Zhang and Xiang(2016)}]{ZhangXiang2016}
Zhang, C., Xiang, Y., 2016. {On the oracle property of adaptive group Lasso in
  high-dimensional linear models}. Statistical Papers 57~(1), 249--265.

\bibitem[{Zhao and Yu(2006)}]{ZhaoYu2006}
Zhao, P., Yu, B., 2006. {On Model Selection Consistency of Lasso}. The Journal
  of Machine Learning Research 7, 2541--2563.

\bibitem[{Zou(2006)}]{Zou2006}
Zou, H., 2006. {The adaptive lasso and its oracle properties}. Journal of the
  American Statistical Association 101~(476), 1418--1429.

\end{thebibliography}

\end{document}